\documentclass[preprint2]{aastex}
\renewcommand\la{\,\raise 0.3ex\hbox{$<$}\kern -0.75em\lower 0.7ex\hbox{$\sim$}\,}
\renewcommand\ga{\,\raise 0.3ex\hbox{$>$}\kern -0.75em\lower 0.7ex\hbox{$\sim$}\,}

\received{April 24, 2000}
\revised {October 5, 2000}
\accepted{December 12, 2000}
\slugcomment{\sl accepted for publication in Astrophysical Journal,
             main journal} 
\shorttitle{Dust evolution in protostellar accretion disks}
\shortauthors{G. Suttner \& H.W. Yorke}

\begin{document}

\title{Early dust evolution in protostellar accretion disks}

\author{Gerhard Suttner\altaffilmark{1,2}}
\email{suttner@astro.uni-wuerzburg.de}
\affil{$^1$Astronomisches Institut der Universit\"at W\"urzburg,
               Am Hubland, 97074 W\"urzburg, Germany}
\and
\author{Harold W. Yorke\altaffilmark{2,1}}
\email{Harold.Yorke@jpl.nasa.gov}
\affil{$^2$Jet Propulsion Laboratory, California Institute of Technology,
               MS 169-506, 
               4800 Oak Grove Drive, Pasadena, CA 91109, USA}

\begin{abstract}

We investigate dust dynamics and evolution during the formation of
a protostellar accretion disk around intermediate mass stars via 2D numerical
simulations. Using three different detailed dust models, compact
spherical particles, fractal BPCA grains, and BCCA grains,
we find that even during the early collapse and the
first $\sim 10^4$~yr of dynamical disk evolution, the initial dust size
distribution is strongly modified. Close to the disk's midplane coagulation
produces dust particles of sizes of several 10\,$\mu$m (for compact spherical
grains) up to several mm (for fluffy BCCA grains), whereas in the vicinity
of the accretion shock front (located several density scale heights above
the disk), large velocity differences inhibit
coagulation. Dust particles larger than about 1\,$\mu$m segregate from
the smaller grains behind the accretion shock. Due to the combined effects of
coagulation and grain segregation the infrared
dust emission is modified. Throughout the accretion disk a MRN dust
distribution provides a poor description of the general dust properties.
Estimates of the consequences of the ``freezing out'' of molecules
in protostellar disks should consider strongly modified grains.
Physical model parameters such as the limiting sticking strength and the grains'
resistivity against shattering are crucial factors determining the degree of
coagulation reached.  In dense regions (e.g. in the mid-plane of the disk)
a steady-state is quickly attained; for the parameters used here the
coagulation time scale for 0.1$\,\mu$m dust particles is
$\sim 1\;{\rm yr}\;(10^{-12}\,{\rm g\; cm}^{-3} / \varrho)$.
High above the equatorial plane coagulation equilibrium is not reached
due to the much lower densities.  Here, the dust size distribution is
affected primarily by differential advection, rather than coagulation.
The influence of grain evolution and grain dynamics on the disk's near
infrared continuum appearance during the disk's formation phase is only
slight, because the most strongly coagulated grains are embedded deep
within the accretion disk.

\end{abstract}


\section{Introduction}

Dust in protostellar envelopes and accretion disks is a major component
of pre-stellar matter, strongly influencing the thermodynamical and gas
dynamical behavior of these young objects as well as their observable
appearance.  Dust provides the seeds for planetesimals, which
in turn evolve into the constituents of a planetary system:
comets, planets, moons, and the debris associated with asteroids
and Kuiper belt objects.
Interstellar dust evolves significantly from its initial state in
recently formed molecular clouds up to the formation
of planetesimals around stars and, as the gas density increases,
it does so at an ever increasing rate.
In molecular clouds it takes several 10$^6$~yr to build up large
fluffy grains by dust coagulation (Ossenkopf 1993; Weidenschilling \&
Ruzmaikina 1994). In the midplane of accretion disks this time scale shrinks
to about 10$^2$~yr due to the high densities there (Mizuno,
Markiewicz, \& V\"olk 1988; Mizuno 1989; Schmitt, Henning, \& Mucha 1997). 

It would be naive to imply, however, that it is only a matter of time
before coagulation produces the first planetesimals of several km in
size.  Other processes affect grain growth and evolution: orbital decay,
shattering, cratering, sputtering, and compacting of amorphous grains,
in addition to adsorption, outgassing, and chemical reprocessing of
molecules.  These processes depend critically on the physical
conditions within the disk and on the grains' relative velocities.
Large compact dust grains ($\ga 10\,\mu$m) can decouple from the gas
and gain large relative velocities to
each other which could prevent coagulation due to a limited sticking strength.
By contrast, large fractal dust grains --- such as Ballistic Cluster Cluster
Agglomeration (BCCA) particles --- are always well coupled to the gas
component; large BCCA particles do not achieve sufficient relative
velocities to coagulate effectively (Schmitt et al. 1997). The
assumption of an ``intermediate'' type of fractal grains, the so called
Ballistic Particle Cluster Agglomeration (BPCA) particles, could alleviate
this problem. Such coagulates possibly couple sufficiently well to the gas to
prevent high relative velocities and behave in the limit of a large number of
constituent grains like compact particles, so that turbulence and systematic
relative velocities can drive coagulation (Ossenkopf 1993).

A high sticking strength is a necessary prerequisite for effective
coagulation up to planetesimal sizes, since turbulent relative velocities can
prevent the coagulation of pre-planetesimal dust grains (Weidenschilling \&
Cuzzi 1993). This could be achieved with ice or frost layers on the colliding
particles' surfaces (Bridges et al. 1996; Supulver et al. 1997).

Dust coagulation leads to important modifications of the protostellar matter.
Turbulence in accretion disks is strongly coupled to the opacity of the medium
which in turn is dependent on the type of dust material and the dust grain
size distribution.  A high degree of coagulation implies significant reduction
of the dust opacity (Mizuno et al. 1988). Reduced opacity is necessary
to damp turbulence which otherwise would be a mayor obstacle for the
pre-planetesimal dust particles to settle down to the disk's equatorial
plane (Weidenschilling 1984). During the formation of an accretion disk,
however, virtually unprocessed dust material is continuously being supplied by
the parent molecular cloud core.  ``Second generation'' coagulation occurs
(Mizuno 1989) which also influences opacity and the dynamics of the disk.

For this study we calculated the dust evolution during the formation of a
protostellar accretion disk using a multi-component radiation
hydrodynamics code, an improved version of a RHD-code designed
for one component (Yorke, Bodenheimer, \& Laughlin 1995;
Yorke \& Bodenheimer 1999).
Different dust models are applied and the influence of the dust--dust sticking
strength is investigated. Brownian motion, turbulence, differential
radiative acceleration and gravitative sedimentation with size
dependent relaxation time scales are
considered as sources of relative velocities between the dust
grains. Mean dust opacities are calculated directly from the actual dust size
distribution and are continuously updated in the radiation transport module.
Finally, a diagnostic frequency and angle dependent radiation transport code
is used to produce synthetic dust continuum emission maps and spectra which
give information on observational consequences.

In section 2 we describe our models for the dust and
in section 3 we sketch our numerical approach to the problem.
The initial conditions for the particular cases considered are
introduced in section 4 and the numerical results of the simulations are
presented in section 5.  In the final section 6 these results are
discussed in light of observable consequences and conclusions are drawn.

\section{Physics of dust grains in star forming regions}

According to our present picture of the interstellar medium,
dust grains more or less uniformly permeate gas clouds,
contributing about one percent of their total mass. The grain size
distribution follows a power law with an exponent of about $-3.5$
(Mathis, Rumpl, \& Nordsieck 1977; hereafter MRN). MRN estimated maximum
and minimum grain sizes at several nm and several $\mu$m,
respectively. The constituent dust grains were assumed to be mainly
``astrophysical'' silicates with some contribution by ``graphites''.
With this dust model they could fit the overall
extinction curve of the diffuse interstellar medium quite well. However,
observations of dense molecular cloud cores show that the upper limiting grain
size is shifted to larger radii (polarization measurements of Vrba, Coyne, \&
Tapia 1993; theoretical predictions by Fischer, Henning, \& Yorke 1994).
Coagulation is assumed to grow large grains in this environment. 

Three different dust models are considered. 
First, we treat the dust as compact spherical particles.  For
this model there are a several theoretical
studies which deal with dust opacities (Yorke 1988), sticking strengths
(Chokshi, Tielens, \& Hollenbach 1993), the physics of
dust shattering (Jones, Tielens, \& Hollenbach 1996),
and gas--dust coupling strengths (Yorke 1979). Thus,
the basic physical properties are reasonably well defined, although
more recent experimental work indicates that the critical sticking
velocities tend to be higher than theoretical estimates by
about a factor of 10 (Poppe \& Blum 1997).

We next consider fractal BPCA grains and BCCA grains. Ossenkopf (1993)
developed a theoretical dust model for these fluffy particles and
provided analytical expressions for the gas--dust
and the dust--dust interaction cross sections as a function of the compact
volume of the grains. Henning \& Stognienko (1996) have calculated
opacities for these fractal dust particles and Wurm (1997) generalized the
sticking strength of Chokshi et al. (1993) to fractal
coagulates. Again, recent experimental studies indicate critical sticking
velocities to be about an order of magnitude larger than the theoretical
values (Poppe \& Blum 1997).

\subsection{Compact spherical grains}\label{SEcompact}

\subsubsection{Opacities and radiative acceleration}

\begin{figure}[tb]
\centerline{\includegraphics*[width=78mm]{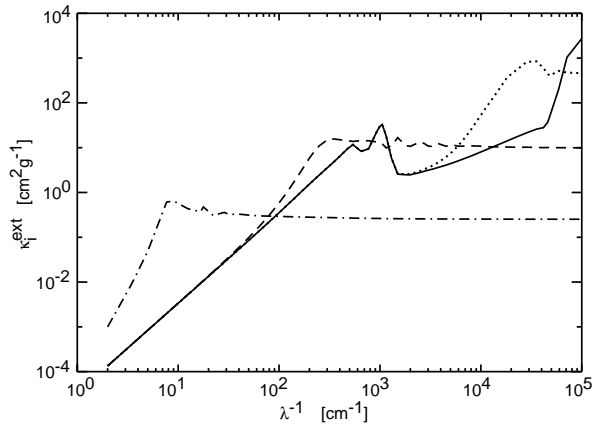}}
\caption{Specific extinction coefficient of compact spherical dust
  grains with sizes of 5\,nm ({\it solid line}), 0.1\,$\mu$m
  ({\it dotted line}), 5\,$\mu$m ({\it dashed line}) and 0.2\,mm
  ({\it dot-dashed line}).}
\label{PIopac1}
\end{figure}

Because dust is the predominant source of extinction
in the temperature, density, and wavelength regimes considered
(e.g. Yorke \& Henning 1994), we have included only the dust's
contribution to opacity and radiative acceleration. The specific
extinction $\kappa^{\rm ext}_{\lambda,i}$ of the component $i$
at wavelength $\lambda$ is proportional to the
specific cross section of the spherical grains:
\begin{eqnarray}
\kappa^{\rm ext}_{\lambda,i}&=&Q^{\rm ext}_{\lambda,i}
    {\rm\pi} a_i^2 / m_i    \\
\kappa^{\rm ext}_\lambda&=&\sum_{i=1}^N \varrho_i
    \kappa^{\rm ext}_{\lambda,i} / \sum_{i=1}^N\varrho_i \\
\kappa^{\rm p}_{\lambda,i}&=&\kappa^{\rm ext}_{\lambda,i}
    \ (1-A_{\lambda,i} g_{\lambda,i})
\end{eqnarray}
The extinction gain factor $Q^{\rm ext}_{\lambda,i}$, the albedo
$A_{\lambda,i}$, and the asymmetry parameter
$g_{\lambda,i}=\left<\cos{\Theta_\lambda}\right>_i$
are calculated from Mie-theory using the dielectric constants for
``astronomical'' silicates given by Draine \& Lee (1984). The net specific
extinction of the dust grains $\kappa^{\rm ext}_\lambda$ is the
weighted sum over
the different constituent grain sizes (Fig. \ref{PIopac1}). Here,
$\kappa^{\rm p}_{\lambda,i}$ is the specific radiation pressure
opacity needed to
determine the interaction of the dust particles with the stellar radiation
flux ${\bf F}_\lambda$. The radiative acceleration of the dust particles is
given by:
\begin{equation}
  \left(\frac{\partial{\bf v_i}}{\partial t}\right)_{\rm rad} =
  \frac 1 {\rm c}
  \int_0^\infty\kappa^{\rm p}_{\lambda,i}{\bf F_\lambda}\, d\lambda
\end{equation}

\subsubsection{Sticking of grains}\label{SEspstick}

Sticking of the spherical dust grains can be described by application of the
theory of elasticity (Chokshi et al. 1993). One
determines the binding energy of two spheres brought together under
consideration of elastic wave dissipation in the two bodies. The resultant
binding energy is set equal to the relative kinetic energy of the two
colliding particles. This gives a critical sticking velocity, above which
coagulation ceases and the particles bounce. Thus, $v_{\rm stick}$ is given by:
\begin{eqnarray}\label{eq-stickE}
  E_{\rm stick} \!\!\!\!&=&\!\!\!\!
    9.6\ \gamma^{5/3}\ {\rm R}^{4/3}\ {\rm E}^{-2/3}        \\
\label{eq-stickv}
  v_{\rm stick} \!\!\!\!&=&\!\!\!\! \sqrt{2\, E_{\rm stick} / \mu}
\end{eqnarray}
 
The constants $\gamma$ and E denote the surface energy per unit area and
Young's modulus of the dust material, respectively.
${\rm R}=r_1 r_2 / (r_1 + r_2)$
is the reduced radius of the contact surfaces and $\mu =m_1 m_2 / (m_1 + m_2)$
is the reduced mass of the two colliding dust particles.
This expression differs by a numerical factor of
about 3 from the original formula by Chokshi et al. (1993) when
applied to spheres, because a correction of Dominik \& Tielens (1997)
has been applied (see discussion by Wurm 1997). Note that Beckwith,
Henning, and Nakagawa (2000) give a slightly different formula for
$v_{\rm stick}$ (quoting the same authors as above):
\begin{equation}\label{eq-stick3}
  v_{\rm stick} = 1.07 \frac{\gamma^{5/6}}{{\rm R}^{5/6}\ {\rm E}^{1/3}\
           \rho_g^{1/2}} \; ,
\end{equation}
where $\rho_g$ is the specific mass of the grain material.  The formulae
\ref{eq-stickE}/\ref{eq-stickv} and Eq. \ref{eq-stick3} are identical
when $r_1=r_2$ and $m_1=m_2$ but differ by 18\% (50\%) when the mass
ratio of the colliding particles is 10 (100) and by a factor of almost
two for extreme dust mass ratios.

An ice layer on the grains' surfaces enhances the critical sticking
velocity by more than an order of magnitude compared to a pure silicate
surface. The ice surfaces are destroyed at dust temperatures of about 125\,K.
Thus, pure silicate grains are assumed where the dust temperature
exceeds the ice
melting limit. According to the experiments of Poppe \& Blum (1997) the
critical sticking velocities are increased by a factor of 10 for the
simulations.

\subsubsection{Grain shattering}

When the relative velocity rises above a critical velocity
$v_{\rm crit}$, the dust
particles are shattered. This is a gradual process, starting with crater
formation on the grains' surfaces and ending with total disruption of
projectile and target. We assume a critical shattering velocity
$v_{\rm crit}$ for silicates (see Jones et al. 1996):
$$
v_{\rm crit} = 2.7\;{\rm km\;s^{-1}}
$$

Jones et al. (1996) also give analytic expressions for the ejected mass during a
shattering encounter. When half of the mass of the target particle (the larger
of two colliding dust grains) is shocked, they assume total disruption;
the debris particles are assumed to follow a power law size distribution
with an exponent of $-3.0$ to $-3.5$. In our simulations we
choose an exponent $-3.5$. The upper size limit of the fragments first grows
with increasing collision velocity (i.e. increasing crater volume) and starts
to decrease inversely proportional to the relative velocity when disruption
starts (see appendix).

\subsubsection{Gas--dust interaction}\label{SEgasdusti}

The dust grains are coupled to the gas by dynamical friction. According to the
gas densities in star forming regions the Epstein coupling law (Epstein 1923;
Weidenschilling 1977) is valid because the mean free path $\lambda_f$ of the
gas molecules is large compared to the radii of the dust particles:
\begin{eqnarray}
\lambda_f&=& 1 / n_{gas}\sigma_{gas}                            \cr
  &\approx& 10^5\,{\rm cm}
  \left[\frac{\varrho}{\rm 10^{-14}\,g\,cm^{-3}}\right]^{-1}\gg a_i
\end{eqnarray}
Here, we use an extension of the Epstein law for superthermal relative
velocities (Yorke 1979), which often occur in later stages of the
protostellar collapse:
\begin{equation}
  \left(\frac{\partial{\bf v_i}}{\partial t}\right)_{\rm ww} \hspace{-1mm}=
  \frac{4}{3}\varrho\frac{\sigma_i}{m_i}
  \sqrt{c_s^2+({\bf v}-{\bf v_i})^2}({\bf v}-{\bf v_i})
\end{equation}
The interaction term $(\partial{\bf v_i} / \partial t)_{\rm ww}$
describes the coupling of dust particles with cross
section $\sigma_i$ and mass $m_i$ with gas of density $\varrho$ and isothermal
sound speed $c_s$.

\subsection{Fractal grains}\label{SEfrac}

\subsubsection{Mass to radius relation}\label{SEfracrad}

There is no simple relation between mass and radius for fractal grains
as there is for compact spherical dust particles. Here, a grain model which
describes the transition from the compact constituent grains (at the lower
radius limit of the size distribution) to the fractals (at the higher end) must
be applied. For BPCA and BCCA grains Henning \& Stognienko (1993) give an
analytic expression for the filling factor $f$ in relation to the extremal
radius $r_{ex,i}$ of the grains:
\begin{eqnarray}
\!\!\!m_i\!\!\!\!&=&\!\!\!\!\frac{4{\rm\pi}}{3}\varrho_{\rm bulk}\
  r_{ex,i}^3\ f_i\\
\!\!\!f_i^{PCA}\!\!\!\!&=&\!\!\!\!0.0457 \left\lbrack1+696
  \left(\frac{r_{ex,i}}{0.01\,{\rm\mu m}}\right)^{\!-3.93}\right\rbrack\\
\!\!\!f_i^{CCA}\!\!\!\!&=&\!\!\!\!0.279 \left(\frac{r_{ex,i}}
  {0.01\,{\rm \mu m}}\right)^{-1.04}\cr
  &&\hspace{1ex}\left\lbrack 1+4.01\left(\frac{r_{ex,i}}
  {0.01\,{\rm \mu m}}\right)^{-1.34}\right\rbrack
\end{eqnarray}
The extremal radius $r_{ex,i}$ is defined as the radius of the minimal
envelope sphere covering a fractal particle with bulk density
$\varrho_{\rm bulk}$.
The constituent grains are assumed to have radii of 0.01\,$\mu$m.

\subsubsection{Opacities and radiative acceleration}

\begin{figure}[tbh]
\centerline{\includegraphics*[width=78mm]{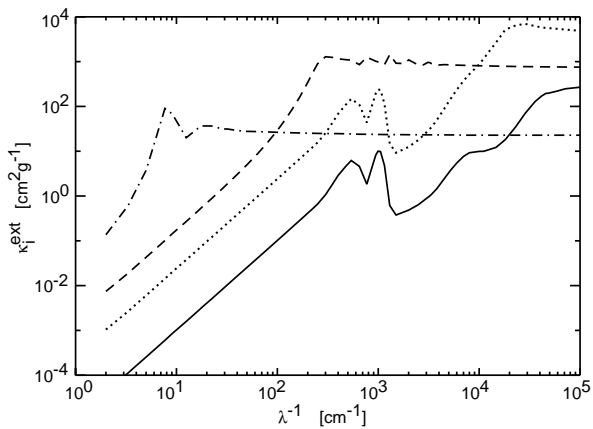}}
\caption{Specific extinction coefficient of fractal dust particles with sizes
  of 5\,nm ({\it solid line}), 0.1\,$\mu$m ({\it dotted line}), 5\,$\mu$m
 ({\it dashed line}), and 0.2\,mm ({\it dot-dashed line}).}
\label{PIopac2}
\end{figure}

Dust opacities for coagulated grains, averaged over a given size
distribution, have been calculated by Henning \&
Stognienko (1996). R. Schr\"apler (private communication), with
the authors' permission kindly provided us with their basic size
dependent specific extinction coefficients.
We used the opacities for olivine
([Fe,Mg]$_2$SiO$_4$) dust grains (Fig. \ref{PIopac2}).

\subsubsection{Sticking of grains}

In contrast to compact spherical grains the fractal coagulates stick at their
limb structures which are built by the small constituent grains. The reduced
radius `R' in Eq. \ref{eq-stickE} for the critical sticking energy must be
calculated using the radii of these constituent grains. In this manner the
formula for the sticking velocity by Chokshi et al. (1993) can
be generalized to fractal particles (see section \ref{SEspstick}).

\subsubsection{Grain shattering}

Two approaches were followed. First, grain shattering is treated the same way
as for the compact spherical grains. The analysis depends only on material
parameters, the relative velocity, and the masses of the colliding particles.
Although this assumption may seem inadequate, it poses an upper limit to the
grains' resistance against shattering. Fractals seem to be rather fragile
constructs with low binding energy when only van der Waals adhesion is
considered. However, experimental studies show that fractal particles are
always more resistant against destruction than the predictions of theoretical
models (Poppe \& Blum 1997; Wurm 1997). Fractals possess a multitude
of vibrational
excitation modes which provide a wide channel to dissipate the kinetic energy
of the impacting grain.

As a second approach we adapt the shattering model of Dominik \& Tielens
(1997):
The critical shattering velocity $v_{\rm crit}$ is proportional to the
sticking velocity $v_{\rm stick}$ and the number of contact points
between the two
colliding grains.  The values thus obtained are rather low compared
to the model of Jones et al. (1996).

\subsubsection{Gas--dust interaction}

The gas--dust interaction also has to be modified due to the fractal
structure of BPCA grains. In order to find an analytic expression
for the effective cross section,
Ossenkopf (1993) fitted numerical calculations of the cross
section $\sigma$ of coagulates in dependence of their compact volume $V$:
\begin{eqnarray}
\frac{\sigma}{\sigma_0}&=&
  a\left(\frac{V}{V_0}\right)^{\frac{2}{3}}
  \exp{\left\lbrack-\frac{b}{(V/V_0)^c}\right\rbrack}\\
a&=&\left\lbrace 15.2,\ \ \  2 \le V/V_0 \le 1000\atop
  4.7,\ \ \ \ \ \ \ \ \ \  V/V_0 > 1000\right.\cr
b&=&\left\lbrace 2.86,\ \ \  2 \le V/V_0 \le 1000\atop
  9.02,\ \ \ \ \ \ \ \ \  V/V_0 > 1000\right.\cr
c&=&\left\lbrace 0.096,\ \  2 \le V/V_0 \le 1000\atop
  0.503,\ \ \ \ \ \ \  V/V_0 > 1000\right.\cr\nonumber
\end{eqnarray}

For BCCA particles the following expression has been derived:
\begin{eqnarray}
\frac{\sigma}{\sigma_0}\!\!\!\!\!&=&\!\!\!\!\!
\cases{
  \raise 1mm\hbox{\rm like\ \ \ BPCA}
 &\raise 1mm\hbox{$\frac{V}{V_0} < 25$}\cr
  \lower 7mm\hbox{$0.692\left(\frac{V}{V_0}\right)^{0.95}
    \left(1+\frac{0.301}{\ln{V/V_0}}\right)$}
 &\lower 7mm\hbox{$\frac{V}{V_0} \ge 25$}\cr
}
\end{eqnarray}
Here, the normalization factors $\sigma_0$ and $V_0$ are the cross section
and the volume of the (compact) constituent grains $\sigma_0=\pi\cdot
(0.01\,\mu{\rm m})^2$ and $V_0=4\pi/3\cdot (0.01\,\mu{\rm m})^3$. The compact
volume can be calculated using the filling factor of section \ref{SEfracrad}.
Thus, a relation between $r_{ex}$ and the cross section $\sigma$ of the
grains can be established and used in the interaction term of section
\ref{SEgasdusti}.

\subsubsection{Dust--dust interaction}

For the dust--dust interaction (coagulation) yet another radius definition has
been introduced. The toothing radius $r_{tooth}$ is defined as half the
average distance of the center of mass of two sticking grains. It measures the
penetration of two fractal grains and has been determined by Ossenkopf (1993)
by fits to numerical calculations which simulated the scan of the ``coastline''
of the larger dust particle by the smaller one. Thus, the collisional cross
section for grain--grain collisions $\sigma_{coll}$ reads:
\begin{eqnarray}
\sigma_{coll}\!\!\!\!&=&\!\!\!\!\sigma^{(1)}\ (1-\zeta^2)            \cr
  &+&\!\!\!\!\left[4\pi-8.3\ (1-\zeta^{-1.22})\right](r_{tooth}^{(2)})^2
\end{eqnarray}
with:
\begin{eqnarray}
 \zeta\!\!\!&=&r\!\!\!_{tooth}^{(2)} / r_{tooth}^{(1)}               \\
 r_{tooth}\!\!\!&=&\!\!\!0.72\ V^{1/3}\ x^{0.21}
              \left(1-\frac{0.216} {x^{1/3}}\right)                  \\
 x\!\!\!&=&\!\!\!\sigma^3 / V^2
\end{eqnarray}
The superscripts $(1)$ and $(2)$ denote the grain with the larger and
the smaller toothing radius, respectively.

\subsection{Coagulation and shattering}\label{SEcoagshat}

Adding the different source terms of grain acceleration, it becomes
obvious that the dust particles will gain relative velocities to each other.
The absolute values of these velocities depend on the dust absorption
and scattering cross sections, the radiative
flux, the gas density and the specific cross section of the grains:
\begin{eqnarray}
  \left(\frac{\partial{\bf v_i}}{\partial t}\right)\!\!&=&\!\!
  \left(\frac{\partial{\bf v_i}}{\partial t}\right)_{\rm grav}\!\!\!+
  \left(\frac{\partial{\bf v_i}}{\partial t}\right)_{\rm rad}\!\!\!+
  \left(\frac{\partial{\bf v_i}}{\partial t}\right)_{\rm ww}       \cr
 \!\!&=& -\nabla\Phi \quad\ \ 
   +\ \ \overline\kappa^{\rm p}_i{\bf F} / c         \cr
 \!\!&+&\!\!\frac{4}{3}\varrho
  \frac{\sigma_i}{m_i}\sqrt{c_s^2+({\bf v}-{\bf v_i})^2}({\bf v}-{\bf v_i})
\end{eqnarray}
In addition to these systematic relative velocities a random contribution
is caused by Brownian motion and turbulence. Because our hydrodynamical grid is
too coarse to resolve typical turbulent length scales, and because turbulence
needs a three dimensional treatment, we apply a turbulence model developed by
V\"olk et al. (1980). Is is coupled to the turbulent angular momentum
transport (Shakura \& Sunyaev 1973) by the parameter $\alpha$. Here, the
macroscopic turbulent velocity $v_{\rm turb}^0$ is set to a fraction $\alpha$ of
the isothermal sound speed $c_s$, and the macroscopic revolution time scale
$t_{\rm turb}^0$ is proportional to the orbital period $\Omega$:
\begin{eqnarray}
v_{\rm turb}^0&=&\alpha\ c_s\\
t_{\rm turb}^0&=&2\pi / \Omega
\end{eqnarray}

A Kolmogorov-type turbulence spectrum is assumed, whereby the
turbulent energy is transported from the largest eddies down to the smallest
at a constant rate. The corresponding scales of the smallest eddies are
defined where the turbulent Reynolds number $Re_s$ of the gas with viscosity
$\eta$ equals one (i.e. the flow becomes laminar, Lang 1974):
\begin{eqnarray}
v_{\rm turb}^s&=&v_{\rm turb}^0\ Re_0^{-\frac{1}{4}}\\
t_{\rm turb}^s&=&t_{\rm turb}^0\ Re_0^{-\frac{1}{2}}
\end{eqnarray}
with:
\begin{equation}
Re_0 = \frac{\varrho\ v_{\rm turb}^0\ \lambda_{\rm turb}^0}{\eta}\approx
  \frac{\varrho\ (v_{\rm turb}^0)^2\ t_{\rm turb}^0}{\eta}
\end{equation}
The back reaction of the gas turbulence on the dust particles depends on
the coupling strength between grains and gas. This strength can be measured by
the correlation time scale $\tau_i$:
\begin{equation}
\tau_i = |{\bf v} - {\bf v_i}| / (\partial v_i/\partial t)_{\rm ww}
\end{equation}

According to the analytical fit of Weidenschilling (1984) to the numerical
results of V\"olk et al. (1980), the random relative velocities between two
grains with correlation time scales $\tau_1$ and $\tau_2$ ($\tau_1$ $\le$
$\tau_2$) can be expressed as follows:
\begin{eqnarray}
\delta v_{\rm turb} = \cases{
 \raise 1mm\hbox{$v_{\rm turb}^s\, \frac{|\tau_1-\tau_2|}{t_{\rm turb}^s}$}
   &\raise 1mm\hbox{if $\tau_1, \tau_2 \le t_{\rm turb}^s$}$\!\!\!\!\!$\cr
 \raise 1mm\hbox{$v_{\rm turb}^0$}
   &\raise 1mm\hbox{if $\tau_1\le t_{\rm turb}^0 \le \tau_2$}$\!\!\!\!\!\!$\cr
 \lower 5mm\hbox{$v_{\rm turb}^0\,
         \frac{t_{\rm turb}^0(\tau_1+\tau_2)}{2\tau_1\tau_2}$}
   &\lower 5mm\hbox{if $t_{\rm turb}^0 \le \tau_1, \tau_2$}$\!\!\!\!$  \cr
 \lower 5mm\hbox{$v_{\rm turb}^0\,
         \frac{3\tau_2}{\tau_1+\tau_2}\sqrt{\frac{\tau_2}{t_{\rm turb}^0}}$}
   &\lower 5mm\hbox{otherwise}                                         \cr
}
\end{eqnarray}
The cutoff of the turbulent eddies at the lower size end of the eddy
spectrum is included according to the considerations of Weidenschilling (1984).
The contribution of Brownian motion to the random part of the relative
velocities between dust grains is only important for small ($a_i < 1\ \mu$m)
grains:
\begin{equation}
\delta v_{\rm brown}=\sqrt{\frac{8 {\rm k} T}\pi\frac{m_i+m_j}{m_i m_j}}
\end{equation}

Thus, the dust particles achieve a total relative velocity
$\delta v_{i,j}=(\delta v_{\rm syst}^2+\delta v_{\rm turb}^2
+\delta v_{\rm brown}^2)^{1/2}$, where
$\delta v_{\rm syst}$ denotes the systematic relative velocities.
These relative
velocities are then evaluated according to the considerations of section
\ref{SEcompact} or \ref{SEfrac}.

If the velocities are sufficiently low, the particles can coagulate.
This is mathematically described by the coagulation equation:
\begin{eqnarray}
\left(\frac{\partial n(m)}{\partial t}\right)_{\rm coag}\!\!\!\!\!\!\!\!
  &=&\!\!\!\frac{1}{2}\int\int\alpha(m^{'},m^{''})\,n(m^{'})\,n(m^{''})\cr
  &&\delta(m-m^{'}-m^{''})\,dm^{'}dm^{''}\cr
  &-&\!\!\!n(m)\int\!\alpha(m,m^{'})\,n(m^{'})\,dm^{'}
\end{eqnarray}
with:
\begin{equation}
\alpha(m^{'},m^{''}) = p\ \sigma_{coll}(m^{'},m^{''})\
  \delta v(m^{'},m^{''})
\end{equation}
The variables $n(m)$, $\sigma_{coll}(m^{'},m^{''})$ and
$\delta v(m^{'},m^{''})$ are the number density, the relative interaction
cross section, and the relative velocity of grains with masses $m$, $m^{'}$ and
$m^{''}$, respectively. The sticking probability $p$ controls the onset of
bouncing when the relative velocities become too high ($p$ $\rightarrow$ 0).
Grain shattering is described by a generalization of the above coagulation
equation:
\begin{eqnarray}
\left(\frac{\partial n(m)}{\partial t}\right)_{\rm shat}\!\!\!\!\!\!\!\!&=&
  \frac{1}{2}\int\int \beta(m^{'},m^{''})\,n(m^{'})\,n(m^{''})      \cr
  &&\hspace{-6mm}
  \gamma(m,m^{'},m^{''}\!\!,\delta v(m^{'}, m^{''}))\,dm^{'}dm^{''} \cr
  &-&\!\!\!n(m)\int\!\beta(m,m^{'})\,n(m^{'})\,dm^{'}
\end{eqnarray}
with:
\begin{equation}
\beta(m^{'},m^{''}) = q\ \sigma_{coll}(m^{'}, m^{''})\
  \delta v(m^{'}, m^{''})
\end{equation}
Here, the function $\gamma$ distributes the shattered dust fragments to the
appropriate mass bins (see appendix). Again, the shattering probability $q$
controls the onset of shattering above the critical velocity ($q$
$\rightarrow$ 1). The total dust evolution operator (coagulation/shattering)
is the sum of both partial operators.

\section{Numerical techniques}

To simulate the collapse of a gravitationally unstable rotating molecular cloud
core we apply a multicomponent radiation hydrodynamics code with detailed
dust dynamics (grain drift, coagulation, shattering). To keep the problem
tractable axial symmetry is assumed (``2.5 D'' in cylindrical coordinates). An
explicit nested grid technique is applied to resolve the inner parts of the
accretion disk (Yorke \& Kaisig 1995).

\subsection{Solution of the coagulation/shattering equation}
\label{SEcoagshatsol}

To solve the combined coagulation/shattering equation numerically the dust size
distribution is binned into $N$ discrete logarithmically spaced  mass intervals.
The continuous equation (section \ref{SEcoagshat}) is therefore discretized:
\begin{eqnarray}\label{eq-dndt}
   \left(\frac{\partial n_k}{\partial t}\right) \!\!\!\!&=&\!\!\!\!
  \left(\frac{\partial n_k}{\partial t}\right)_{\rm coag}
  +\left(\frac{\partial n_k}{\partial t}\right)_{\rm shat}\cr
  \!\!\!\!&=&\!\!\!\!
  \frac{1}{2}\sum_{i=1}^N\sum_{j=1}^N(\alpha_{ij}d_{ijk}+\beta_{ij}g_{ijk})
  n_i n_j \cr
  &&\!\!\!\! -n_k\sum_{i=1}^N(\alpha_{ik}+\beta_{ik})n_i
\end{eqnarray}
with:
\begin{eqnarray}
&&d_{ijk} = \cases{
  \frac{m_i+m_j}{m_k} &if $m_i+m_j \in [m_k^-,m_k^+]$ \cr
  0                   &otherwise                      \cr
}                                                                      \cr
&&m_k^- = (m_k + m_{k-1})/2 \quad m_k^+ = m_{k+1}^-                    \cr
&&g_{ijk} = \frac{m_i+m_j}{m_k}\, G_k(m_i,m_j,\delta v_{ij})           \\
&&G_k(m_i,m_j,\delta v_{ij}) \in [0,1]                                 \cr
&&\sum_{k=1}^N G_k(m_i,m_j,\delta v_{ij})=1                     \nonumber
\end{eqnarray}

The distribution of the shattered fragments is given by the discrete
distribution function $G_k(m_i,m_j,\delta v_{ij})$ (see appendix).
The kernels $\alpha_{ij}$ and $\beta_{ij}$ are the discretized counterparts
of $\alpha(m^{'},m^{''})$ and $\beta(m^{'},m^{''})$ (section \ref{SEcoagshat}).
Substituting backward time differences for all time derivatives,
this nonlinear integro-differential-equation can be brought into the form:
\begin{equation}\label{eq-DnDt}
  ({\bf n}^*-{\bf n})/\Delta t={\cal A}({\bf n}^*){\bf n}^*,
\end{equation}
where ${\bf n}=(n_1,n_2,\ .\ .\ .\ ,n_N)$ are the (known) densities at the
beginning of the time step and ${\bf n}^*$ are the corresponding values
after time $\Delta t$ which are to be determined.
${\cal A}({\bf n}^*)$ is a matrix which is constructed from the right
hand side of Eq. \ref{eq-dndt}.  The advantages of backward time derivatives
are: a) the difference equations are numerically stable for all choices of
time steps $\Delta t$, and b) the numerical solution approaches
the correct steady-state solution as $\Delta t \rightarrow \infty$.
We iteratively solve the implicit equation \ref{eq-DnDt} for each cell
in the numerical grid using a multidimensional Newton-Raphson algorithm.
In order to optimize convergence we adaptively reduce the time step
with respect to the relatively large time step used for the explicit
hydrodynamics. We tested our solver by calculating the solution of simple
coagulation problems for which analytic solutions exist (Wetherill 1990).
The correspondence was very good at high resolutions of mass binning.
At resolutions comparable to those used in the collapse calculations
the accuracy suffers.  Sharply peaked mass distributions and
discontinuous mass distributions become more diffuse with the passage
of time.  We do not consider this to be a serious flaw, however,
because of the large uncertainties of the details of grain growth
and destruction.  The total mass of the dust component was
always conserved within rounding errors (Suttner, Yorke, \& Lin 1999).

\subsection{The multicomponent radiation hydrodynamic code}

An explicit/implicit method is used to solve the coupled hydrodynamic
equations and the equations of radiation transport. The system is separated
applying operator splitting, and the partial equations are then
discretized explicitly or implicitly according to stability considerations.
The dust size distribution is binned into $N$ mass intervals and the
hydrodynamic equations for mass and momentum conservation
are computed for the gas component and for each binned
dust component (grain size) simultaneously.
The equations of mass conservation for gas ($\varrho$)
and dust particles ($\varrho_k$) can be written:
\begin{eqnarray*}
&&\hspace{-7mm}
  \frac{\partial\varrho}{\partial t}+ \nabla\cdot(\varrho{\bf v})=0\\
&&\hspace{-7mm}
  \frac{\partial\varrho_k}{\partial t}+ \nabla\cdot(\varrho_k {\bf v})=
  -\varrho_k\sum_{i=1}^N(\alpha_{ik}+\beta_{ik})\frac{\varrho_i}{m_i} \\
&&+\frac{1}{2}\sum_{i=1}^N\sum_{j=1}^N(\alpha_{ij}d_{ijk}+\beta_{ij}g_{ijk})
  \frac{\varrho_i}{m_i}\frac{\varrho_j}{m_j}\cdot m_k\\
\end{eqnarray*}
Here, dust coagulation and shattering has been included in the equation of
continuity for the dust particles as a source/sink term.

The equations for momentum conservation in cylindrical
coordinates for gas and dust grains are:
\begin{eqnarray*}
&&\hspace{-5mm}
  \frac{\partial(\varrho v_z)}{\partial t}+ \nabla\cdot(\varrho v_z{\bf v})=
  -\frac{\partial p}{\partial z}-\varrho\frac{\partial \Phi}{\partial z} \\
&&\hspace{6mm}+\varrho\sum_{k=1}^N I_k (v_{k,z}-v_z)-( \nabla\cdot
  {\cal Q}_{\rm visc})_z\\
&&\hspace{-5mm}
  \frac{\partial(\varrho v_r)}{\partial t}+ \nabla\cdot(\varrho v_r{\bf v})=
  -\frac{\partial p}{\partial r}-\varrho\frac{\partial\Phi}{\partial r}\\
&&\hspace{6mm}+\varrho\sum_{k=1}^N I_k (v_{k,r}-v_r)+\varrho
  \frac{v_\phi^2}{r}-( \nabla\cdot {\cal Q}_{\rm visc})_r\\
&&\hspace{-5mm}
  \frac{\partial(\varrho v_\phi r)}{\partial t}+ \nabla\cdot(\varrho v_\phi r
  {\bf v})=\\
&&\hspace{6mm}\varrho\sum_{k=1}^N I_k (v_{k,\phi}-v_\phi)\ r
  -( \nabla\cdot {\cal Q}_{\rm visc})_\phi\ r\\
&&\hspace{-5mm}
  \frac{\partial(\varrho_k v_{k,z})}{\partial t}+ \nabla\cdot(\varrho_k
  v_{k,z}{\bf v}_k)=\varrho_k\frac{\overline\kappa^{\rm p}_k F_z}{\rm c}
  -\varrho_k \frac{\partial \Phi}{\partial z}\\
&&\hspace{6mm}-\varrho I_k (v_{k,z}-v_z)\\
&&\hspace{-5mm}
  \frac{\partial(\varrho_k v_{k,r})}{\partial t}+ \nabla\cdot(\varrho_k
  v_{k,r}{\bf v}_k)=\varrho_k\frac{\overline\kappa^{\rm p}_k F_r}{\rm c}
 -\varrho_k \frac{\partial \Phi}{\partial r}\\
&&\hspace{6mm}-\varrho I_k (v_{k,r}-v_r)
  +\varrho_k\frac{v_{k,\phi}^2}{r}\\
&&\hspace{-5mm}
  \frac{\partial(\varrho_k v_{k,\phi}r)}{\partial t}+ \nabla\cdot
  (\varrho_k v_{k,\phi}r{\bf v}_k)=\\
&&\hspace{6mm}-\varrho I_k (v_{k,\phi}-v_\phi)\ r
\end{eqnarray*}
with the interaction term:
\begin{equation}
I_k = \frac{4}{3}\varrho_k\frac{\sigma_k}{m_k}\sqrt{c_s^2+({\bf v}-{\bf v}_k)^2}
\end{equation}

The advection part of these equations is solved by an explicit second
order scheme. The gas--dust interaction terms need an implicit treatment
because of the stiffness of the problem.

The tensor ${\cal Q}_{\rm visc}$ which appears in several of the above
equations denotes the viscous stress tensor of the $\alpha$-viscosity
(Shakura \& Sunyaev 1973):
\begin{equation}
  {\cal Q}_{\rm visc} = \varrho \nu \, {\bf e} \; ,
\end{equation}
where ${\bf e}$ is the shear tensor.  $\nu$ is calculated from
\begin{equation}
  \nu = \alpha c_s H \approx 0.7\, \alpha c_s^2 / \Omega \; ,
\end{equation}
where the density scale height $H$ has been replaced by an expression
valid for equilibrium ``thin'' disks.   For the calculation described
in this study both sound speed $c_s$
and angular velocity $\Omega$ are approximately constant along 
the surfaces of cylinders within the equilibrium disk (c.f the
theoretical discussion in Tassoul 1978).  Thus, $\nu$ varies principally
as a function of the radial distance within the disk.

The viscosity as described above is applied to the entire grid.
Within the accretion disk
it modifies the flow by allowing angular momentum
to be transfered radially outwards within the disk.  The parameter
$\alpha$ is continuously adjusted according to the Toomre stability
criterion and is allowed to vary within the range of $10^{-3}$
to 0.1 (c.f. Yorke \& Bodenheimer 1999).  We define the Toomre
parameter $Q_T$ within the accretion disk for each time step
by the minimum value of:
\begin{equation}
  Q_T = \hspace{1mm} {\rm min}
     \hspace{-7mm}\lower 2.4mm\hbox to 11mm{\scriptsize
     $r\! \le\! R_{\rm disk}$} \hspace{1mm}
     \left[\; \Omega c_s / \pi G \Sigma\; \right]_{z=0} \; ,
\end{equation}
where $\Sigma(r) = \int \varrho dz$ is the disk's surface density.
If $Q_T$ drops below 1.3 (i.e. nonradial instabilities can be expected
to occur), we increase $\alpha$ by a small factor (typically 1.002),
if necessary to its maximum value 0.1. If $Q_T$ increases above 1.5
(the disk becomes stable), $\alpha$ is reduced by a small factor
(typically 0.999).  Generally, $Q_T$ `hovers' at
either 1.3 or 1.5 and $\alpha$ levels off at values somewhere between
$\alpha \approx 0.01$ (for $M = 1$~M$_\odot$) and 
$\alpha \approx 0.08$ (for $M = 10$~M$_\odot$).

Radiation transport is calculated within the framework of the grey
flux limited diffusion approximation (Levermore \& Pomraning 1981;
Yorke \& Bodenheimer 1999):
\begin{equation}\label{eq-FLD}
\frac{\partial a T_d^4}{\partial t} =
   \nabla\cdot \left( \frac{{\rm c}\lambda_R \; \nabla a T_d^4 }
  {\sum_{k=1}^N\overline\kappa_k(T_d)\varrho_k} \right)
   + L_*\delta(V_1) = 0
\end{equation}
with flux limiter $\lambda_R$, Rosseland mean opacity
$\overline\kappa_k(T_d)$, the radiation constant $a$,
and luminosity of the central source $L_*$ (treated as an additional
source term within the volume $V_1$ of the innermost grid cell):
\begin{eqnarray}
\delta(V_1) \!\!\!\!&=&\!\!\!\! \cases{ 1/V_1 &if innermost cell \cr
                                         0    &otherwise}  \\
\lambda_R \!\!\!\!&=&\!\!\!\!
  \frac{1}{\xi}\left(\coth(\xi)-\frac{1}{\xi}\right)\\
  \xi \!\!\!\!&=&\!\!\!\! \frac{| \nabla T_d^4 |}
  {T_d^4 \sum_{k=1}^N\overline\kappa^{\rm ext}_k(T_d)\varrho_k}\\
  \frac{1}{\overline\kappa_k(T)} \!\!\!\!&=&\!\!\!\!
  \int_0^\infty\frac{1} {\kappa^{\rm ext}_{\lambda,k}}
  \frac{{d \rm B_\lambda}}{d T}d\lambda\left/
  \int_0^\infty\frac{{d \rm B_\lambda}}{d T}d\lambda\right.
\end{eqnarray}
Here, ${\rm B}_\lambda ={\rm B}_\lambda(T)$ is the Planck function.
Because we are considering grey radiation transfer only,
the equilibrium temperature of each dust grain is identical to the
radiation temperature $T_d$.  To solve equation \ref{eq-FLD} for
$T_d$ an implicit ADI procedure is used (Douglas \& Rachford 1956).

By contrast, the gas is poorly coupled to the radiation field due
to its low opacity.  The dust grains interact with
the gas by inelastic collisions which contribute a cooling term $\Lambda$ to
the equation for the internal energy density of the gas. The dissipation of
viscous energy leads to an additional gas heating term
${\cal Q}_{\rm visc}:\nabla {\bf v}$
\begin{eqnarray}
\frac{\partial\epsilon}{\partial t}+ \nabla\cdot(\epsilon{\bf v})
  \!\!\!\!&=&\!\!\!\!
  -p \nabla\cdot{\bf v}-\Lambda(\varrho,\varrho_k,T,T_d)        \cr
  &&\!\!\!\! + {\cal Q}_{\rm visc}:\nabla {\bf v}
\end{eqnarray}
Assuming an energy transfer of k$(T_d-T)$ per collision
of a gas molecule with a dust grain the cooling function becomes
($\mu_0$ is mean molecular weight of the gas):
\begin{eqnarray}
 \Lambda(\varrho,\varrho_k,T,T_d) = \hspace{39mm}               \cr
  \frac{\varrho}{\mu_0 m_H}\sum_{k=1}^N
  \frac{\varrho_k}{m_k}\sigma_k\sqrt{\frac
  {8{\rm k}T}{\pi\mu_0 m_H}} {\rm k}(T_d-T)
\end{eqnarray}

The equation of state $p(\varrho,T)$ and the internal energy
$\epsilon(\varrho,T)$ for the gas component
assume molecular gas and includes dissociation of the H$_2$ molecules above
$\approx$ 2000\,K (Black \& Bodenheimer 1975).

Finally, the gravitation potential of the molecular cloud is calculated by
a solution of the Poisson equation, again using ADI:
\begin{equation}
  \Delta\Phi = 4\pi G\ (\varrho+\sum_{k=1}^N\varrho_k) \; .
\end{equation}

\section{Initial and boundary conditions}

\begin{table}[th]
\begin{center}
\caption{INITIAL CONDITIONS}
\hspace{0mm}\\
  \begin{tabular}{ccccc}
    \tableline\tableline\\[-3mm]
          &     &$M_{\rm c}$       &$\Omega$              &$t_{\rm ff}$ \\
    Model &Dust &$[{\rm M}_\odot]$
   &\hspace{-2mm}$[10^{-12}\, {\rm s}^{-1}]$\hspace{-2mm}&$[yr]$ \\[1mm]
    \tableline\\[-3mm]
           1MS &comp. & 1 &1 &8635\\
           3MS &comp. & 3 &3 &4985\\
           5MS &comp. & 5 &4 &3870\\
          10MS &comp. &10 &5 &2730\\
      1MS\_PCA &BPCA  & 1 &1 &8635\\
      3MS\_PCA &BPCA  & 3 &3 &4985\\
      5MS\_PCA &BPCA  & 5 &4 &3860\\
     10MS\_PCA &BPCA  &10 &5 &2730\\
      1MS\_CCA &BCCA  & 1 &1 &8635\\
      3MS\_CCA &BCCA  & 3 &3 &4985\\
      5MS\_CCA &BCCA  & 5 &4 &3860\\
     10MS\_CCA &BCCA  &10 &5 &2730\\[-2mm]
  \end{tabular}
  \tablenotetext{}{{\sc Note.}---
    $M_{\rm c}$: Mass of cloud core; $\Omega$: Angular velocity;
    $t_{\rm ff}$: Free-fall timescale.}
\label{TAinitsimtab}
\end{center}
\end{table}

We start the numerical simulations with an isothermal, uniformly 
rotating molecular
cloud core with a total mass of $1\,$M$_\odot$ to $10\,$M$_\odot$, a radius of
$r=2\times 10^{16}$\,cm and a temperature of $T=20$\,K. This gives an initial
free-fall timescale $t_{\rm ff}$ between $\approx 8600$~yr and
$\approx 2700$~yr for these configurations. The angular velocities $\Omega$
range between $10^{-12}$\,s$^{-1}$ and $5 \times 10^{-12}$\,s$^{-1}$
(see Table \ref{TAinitsimtab}). We consider centrally peaked mass
distributions $\varrho \propto 1/(r^2 + z^2)$.
The total mass contribution of the dust grains is set to
a fraction of $0.25\times 10^{-2}$ of the gas mass (corresponding to the mass
contribution of silicates).

At the outer boundary of the space integration domain
($\sqrt{r^2+z^2}=2\times 10^{16}$\,cm) the
hydrodynamic variables are held constant (no inflow or outflow).
This corresponds to an assumption that no material from the
parent cloud will enter into the portion undergoing collapse.
This does not mean that the mass influx rate of dusty material
onto the new formed disk is suddenly cut off after one free-fall
time.  Instead, it steadily decreases as the material in the outer
zones, initially slowed by pressure gradients in the density-peaked
distribution, is depleted (see Yorke \& Bodenheimer 1999).
Even after three free-fall times, there
is an appreciable mass influx onto the disk.
From other studies (e.g. Mizuno et al. 1988) we know that if
the enshrouding molecular cloud can continuously supply small
dust grains, the grain size distribution will be affected.  This
effect will be present to some extent in the studies discussed here.
The issue of the time dependency of the mass influx is a non-trivial
one, however, and is beyond the scope of the present investigation.

The particle
sizes are distributed according to a MRN power law with an exponent of $-3.5$
which can be transformed to a bin mass distribution ${\rm m_{bin}}(m_k)$:
\begin{eqnarray}
n(a) \!\!\!\!&\propto&\!\!\!\! a^{-3.5}                                    \\
{\rm m_{bin}}(m)d\,{\rm ln}\,m \!\!\!\!&\propto&\!\!\!\! m^{1/6}d\,{\rm ln}\,m
\end{eqnarray}
with grain mass $m$. Figure \ref{dPIspheres} (upper left panel) shows the
initial MRN dust mass distribution. On a logarithmic scale the dust
mass increases with increasing bin mass. The grain sizes range from 5\,nm to
5\,$\mu$m at the beginning. Above 5\,$\mu$m the particle density falls off
proportional to ${\rm m_{bin}}^{-1}$ (this corresponds to $n(a)\propto a^{-7}$).
The dynamics of these large grains at the upper end of the computed size
distribution can be followed throughout the simulation without being relevant
for coagulation (provided that the upper end is chosen far enough away of the
largest particles produced by coagulation). The integration domain in dust
size space ranges from 5\,nm up to 0.2\,mm which spans about 14 orders of
magnitude in grain mass. Whenever the dust temperature is low enough to allow
an ice coating on the grains' surfaces, the sticking propabilities are
modified accordingly.

The innermost cell of the finest grid contains the central protostellar
source and requires special treatment. Its total luminosity $L_*$ can be
approximated by the sum of the core's intrinsic luminosity and the
luminosity due to accretion of material onto the core:
\begin{equation}
L_* = 3L_\odot\left(\frac{M_*}{M_\odot}\right)^3+\frac{3}{4}
  \frac{GM_*\dot M_*}{R_*}
\end{equation}
The radius $R_*$ of the central object is held constant at
10\,$R_\odot$.  $\dot M_*(t)$ and $M_*(t) = \int \dot M_*dt$ result from
the calculations.

\begin{figure*}
\hbox to \textwidth {\includegraphics*[width=80mm]{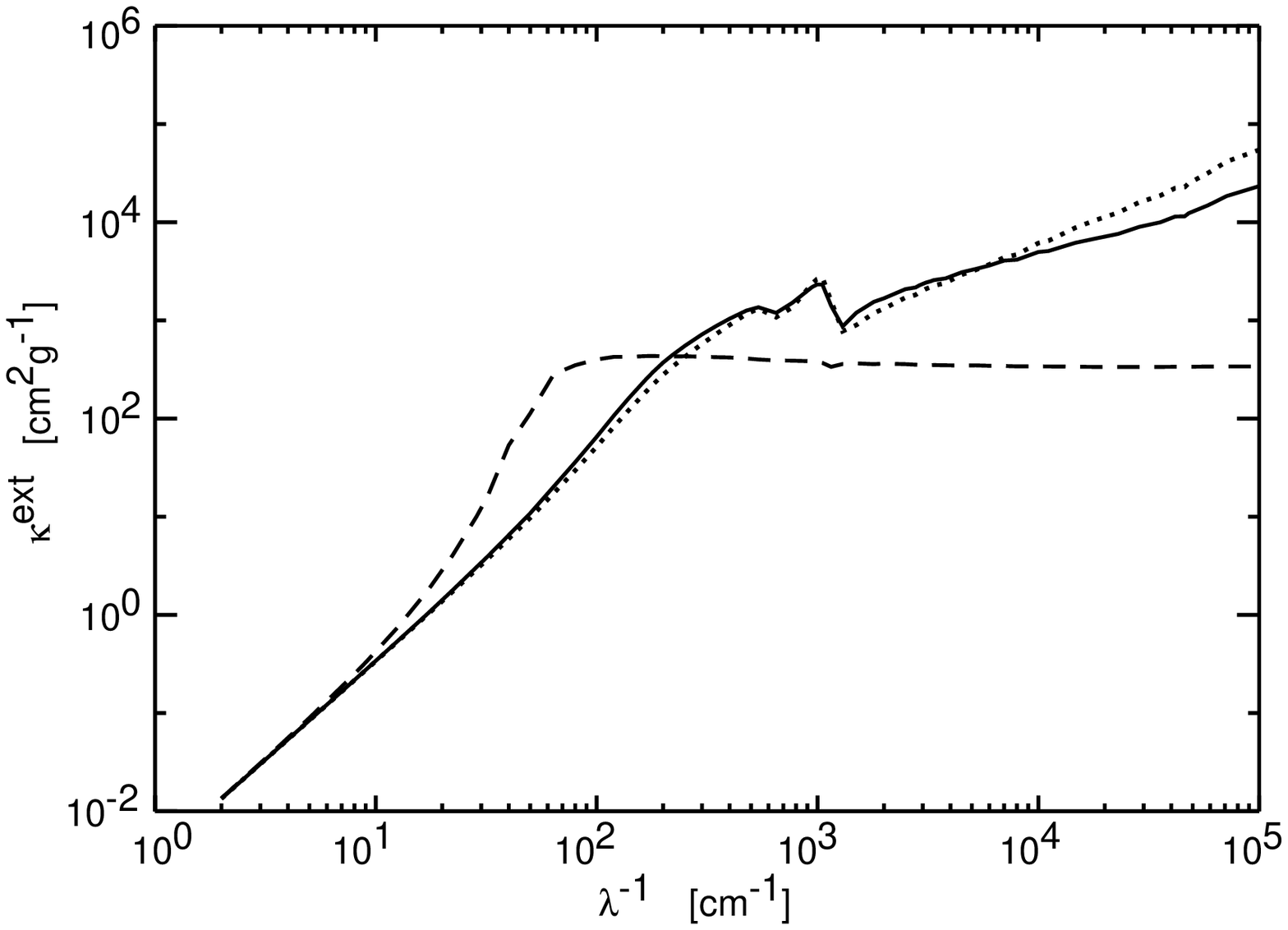}
             \hfil   \includegraphics*[width=80mm]{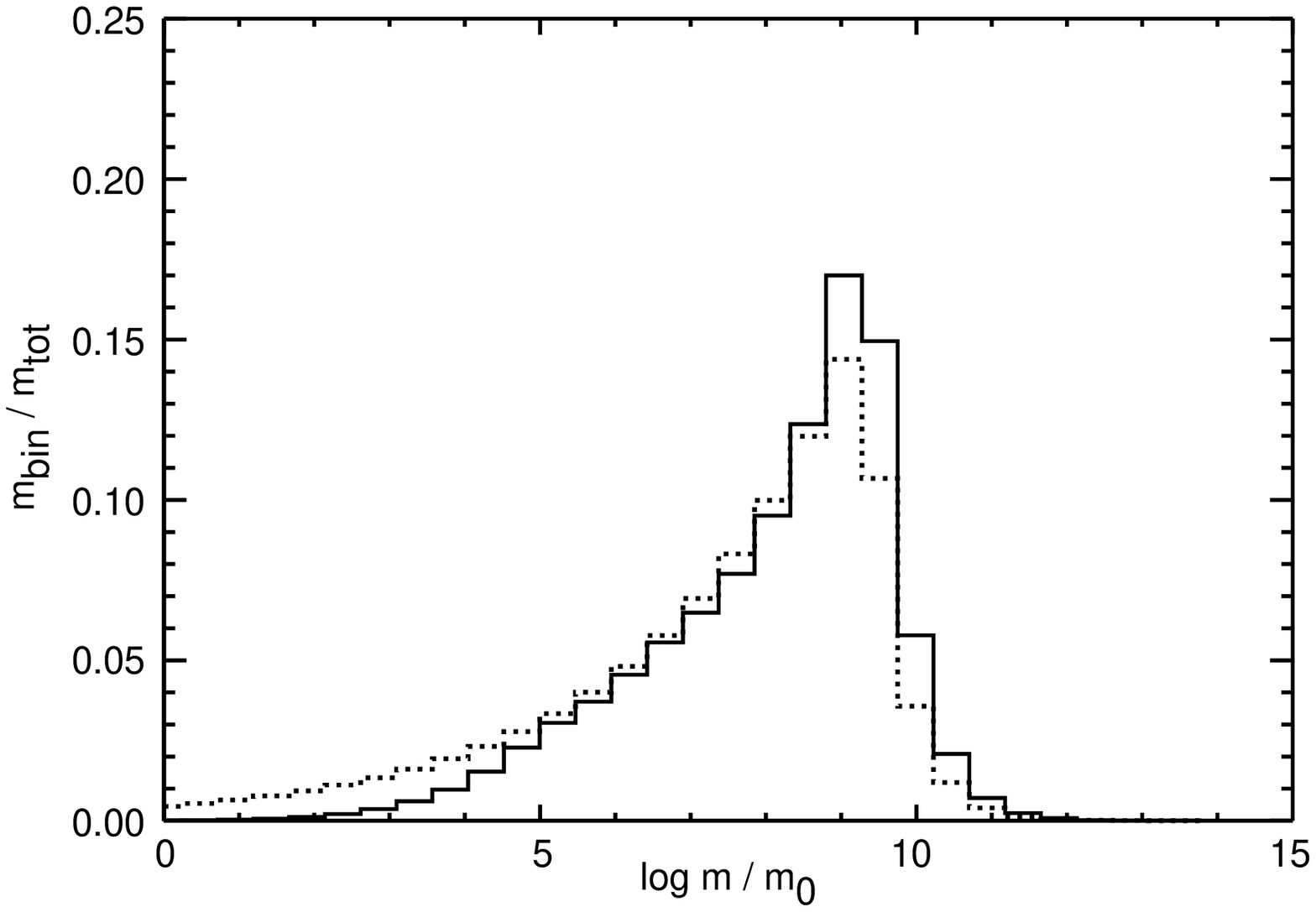}}
\caption{Model 3MS, 11400~yr. Left: The specific extinction
  coefficient of the dust in the accretion shock ({\it dotted
  line}) and in the equatorial plane of the accretion disk
  ({\it dashed line}) is compared 
  with an MRN mass distribution ({\it solid line}).
  Right: Overall mass spectrum at the end of a comparison simulation
  whereby the contribution of
  systematic velocities to the relative motion of the dust grains is
  neglected.}
\label{PIevext}
\end{figure*}

Figure \ref{PIevext} (left panel, {\it solid line}) displays the net specific
extinction coefficient for a gas--dust mixture with compact spherical grains.
The opacities are calculated for the MRN mass distribution shown in Figure
\ref{dPIspheres} (upper left panel). Because the dust opacities are calculated
using the actual grain size distribution they vary as a function of
time and location during the simulation.

\section{Numerical Simulations}

\begin{figure*}
\hbox to \textwidth {\includegraphics*[width=80mm]{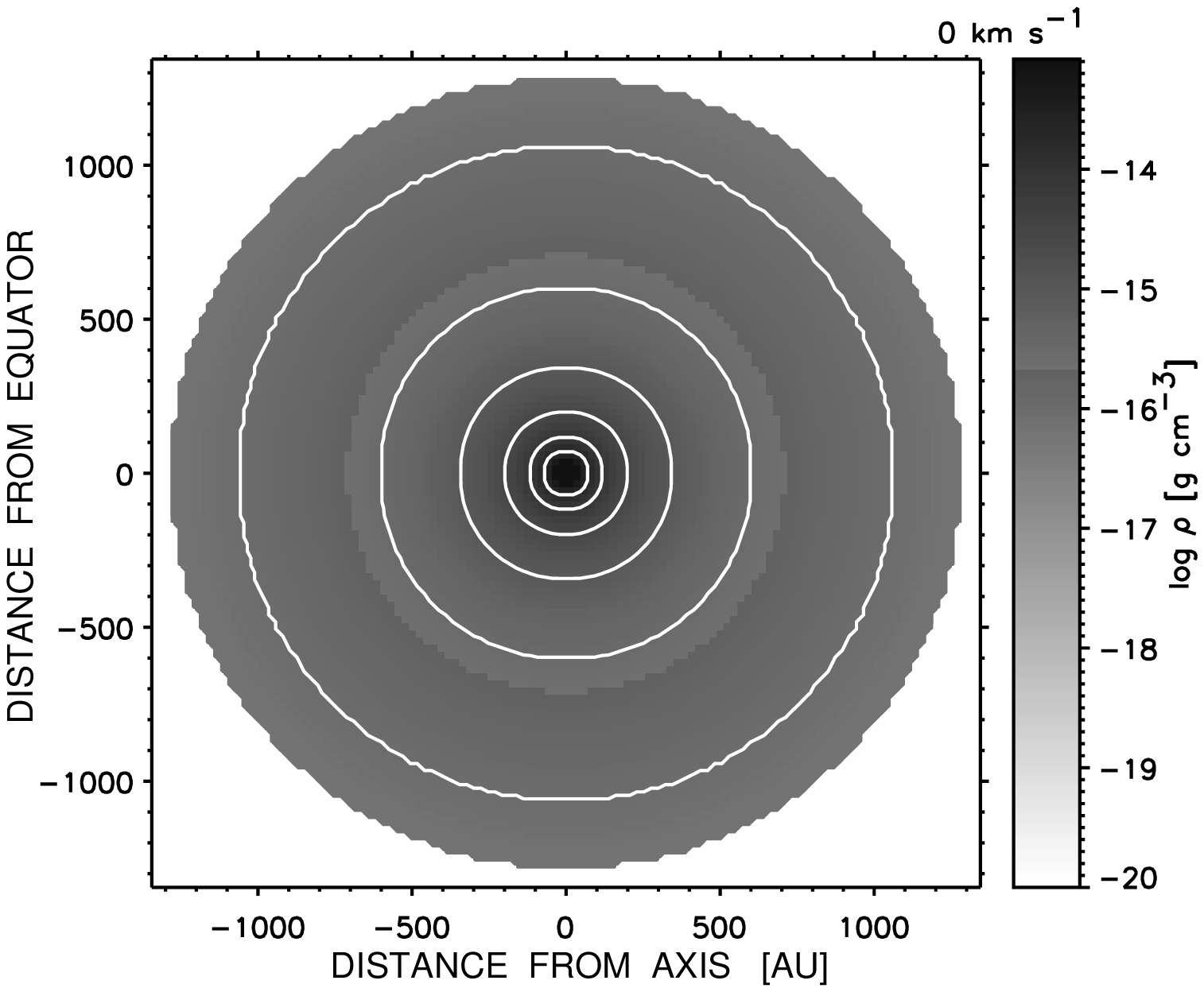}
             \hfil   \includegraphics*[width=80mm]{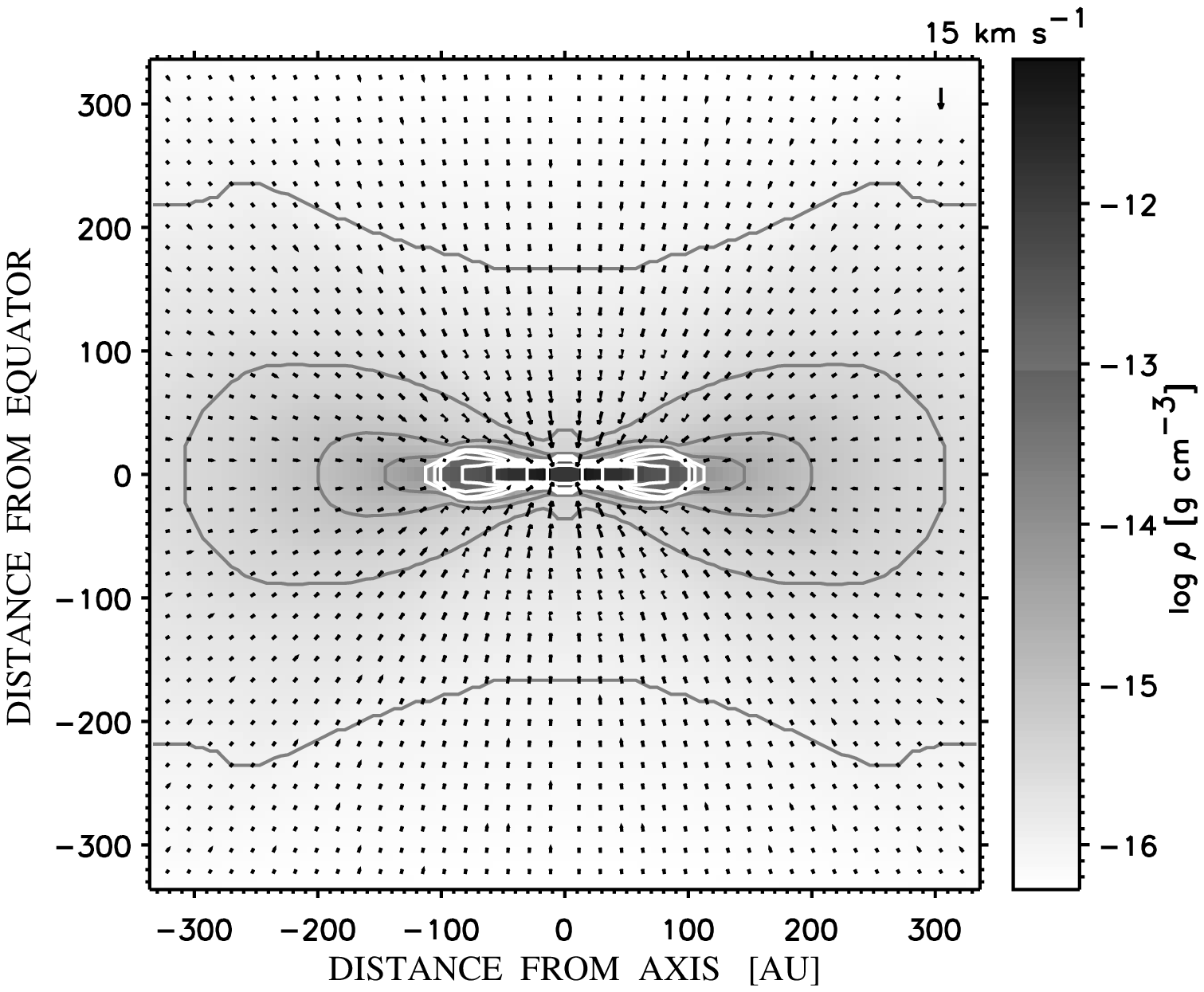}}
\hbox to \textwidth {\includegraphics*[width=80mm]{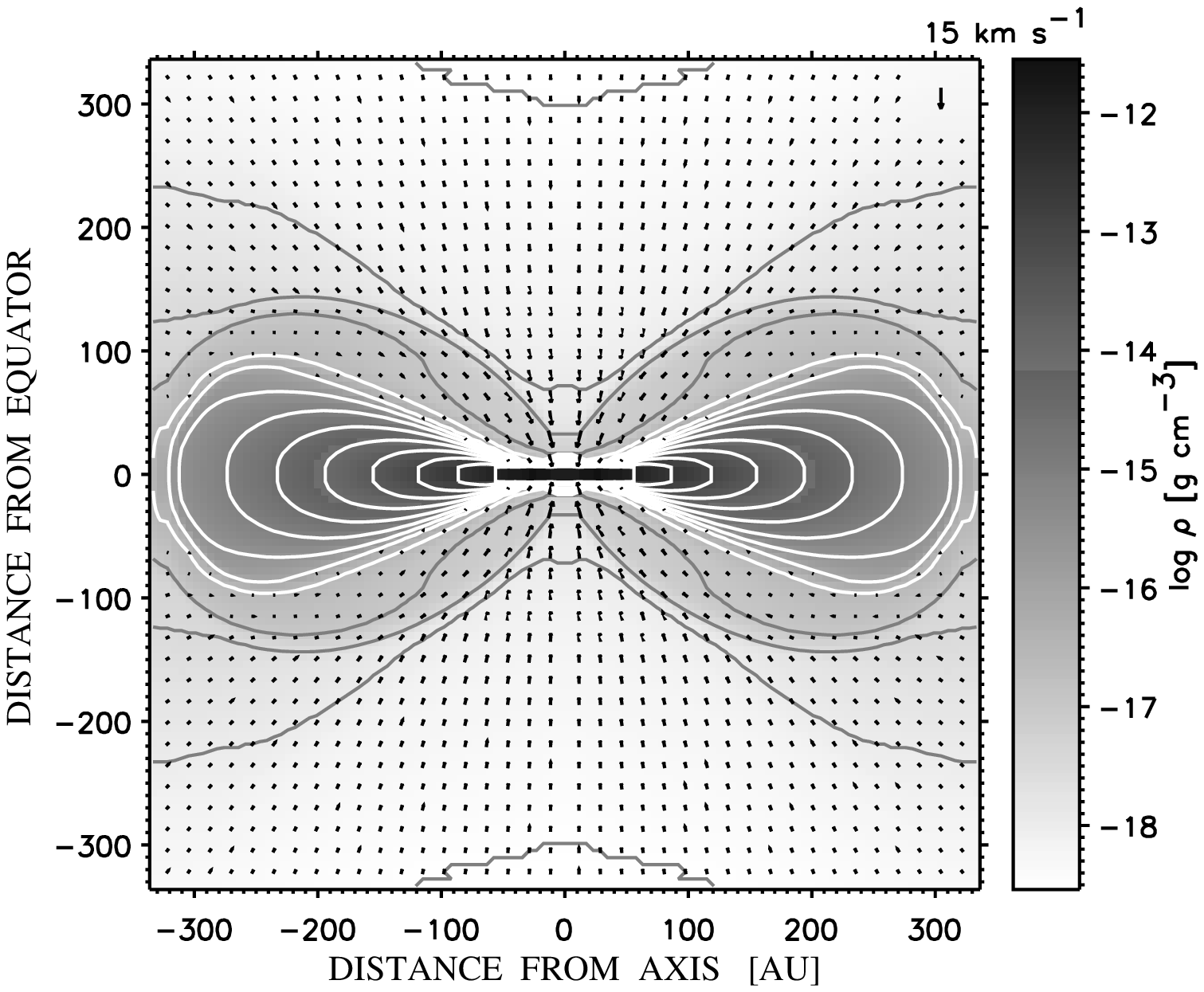}
             \hfil   \includegraphics*[width=80mm]{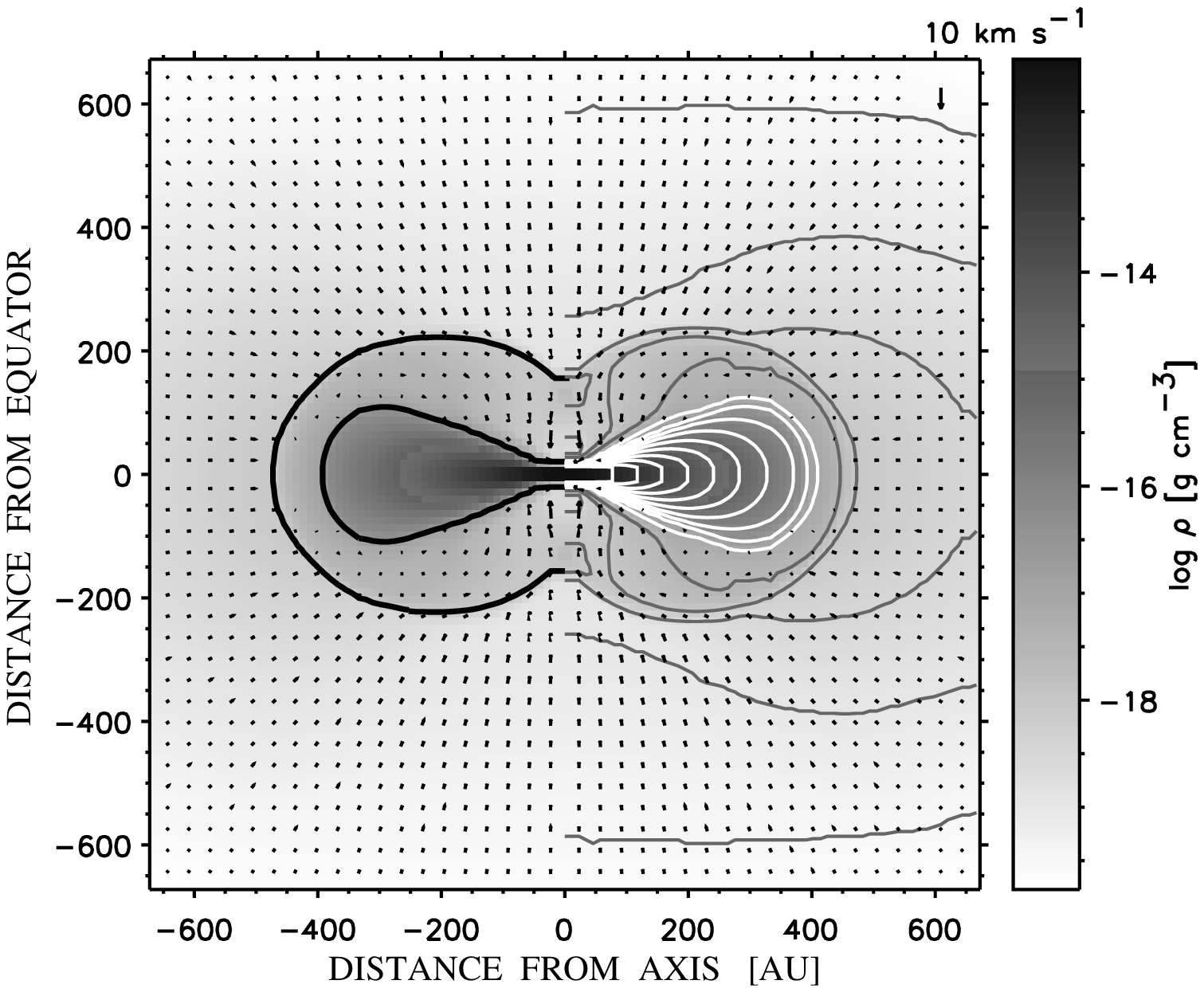}} 
\caption{Evolution of density and velocity
  using the compact sphere dust model 3MS.
  Density contour lines are separated by 
  $\Delta$log $\varrho=0.5$. From left to right and top to bottom
  the following times are shown: 0~yr, 5100~yr, 10300~yr and 11400~yr.
  In the lower right frame the locations of the accretion shocks are
  given by thick dark lines on the left half, where the density contour
  lines (shown on the right half only) begin to tightly bunch.  The
  inner accretion shock encompasses the equilibrium accretion disk,
  where the $(v_r,v_z)$ components of velocity are negligible.
  The outer accretion shock
  is less apparant, because the bunching of a mere two contour lines
  --- corresponding to a jump in density by a factor of $\sim$10
  --- is less conspicuous.}
\label{sPIspheres}
\end{figure*}

\begin{figure*}
\hbox to \textwidth {\includegraphics*[width=80mm]{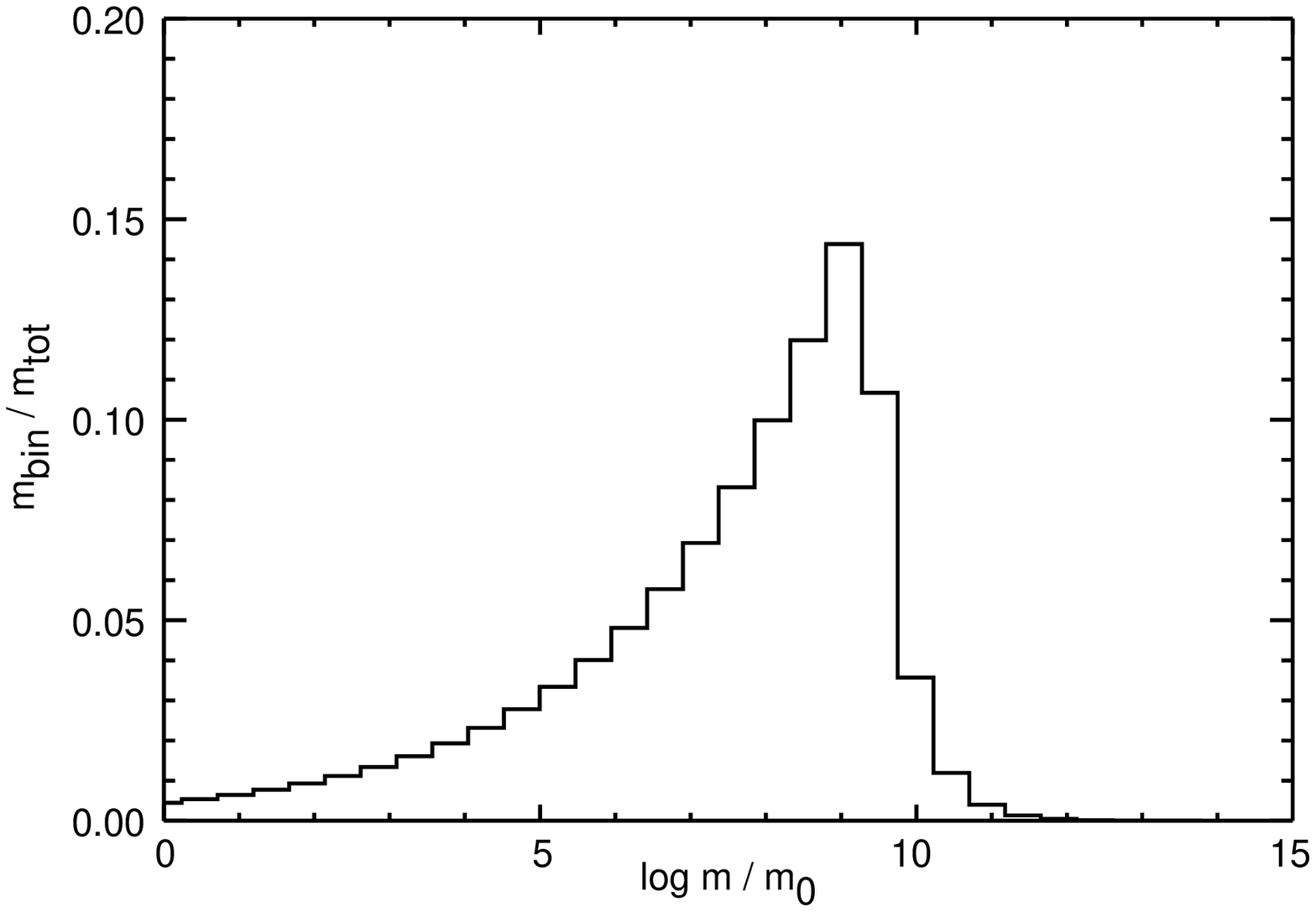}
             \hfil   \includegraphics*[width=80mm]{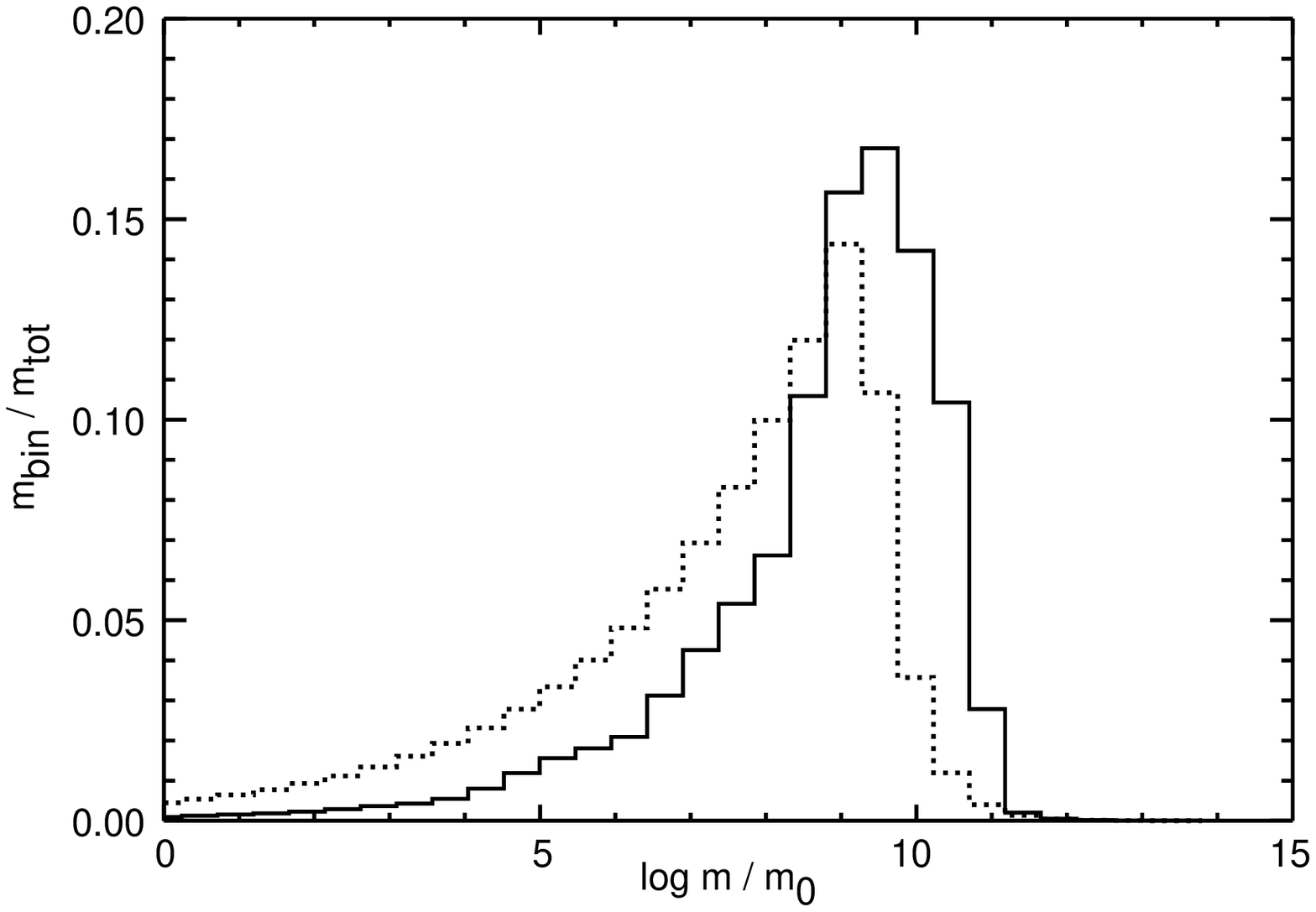}} 
\hbox to \textwidth {\includegraphics*[width=80mm]{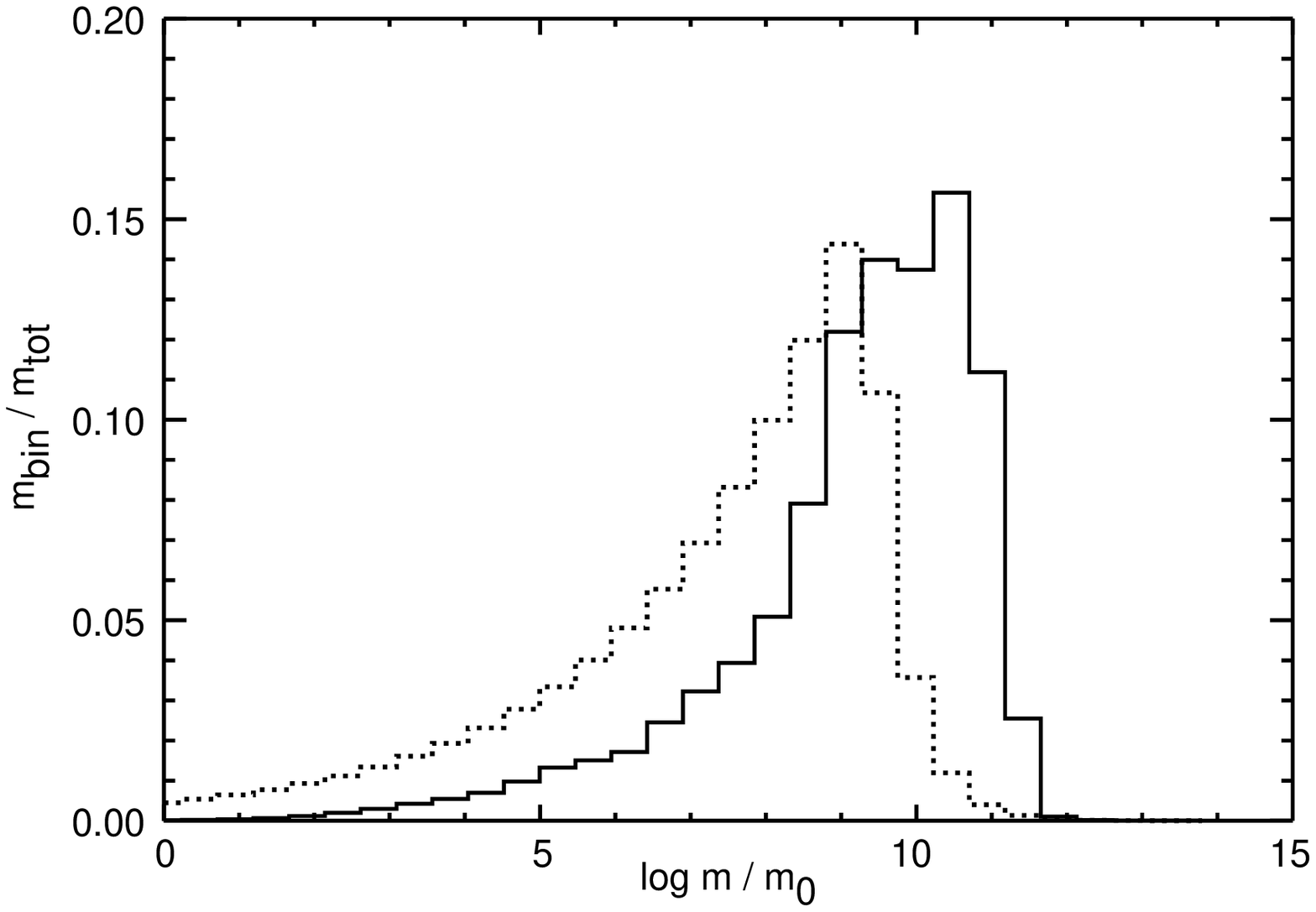}
             \hfil   \includegraphics*[width=80mm]{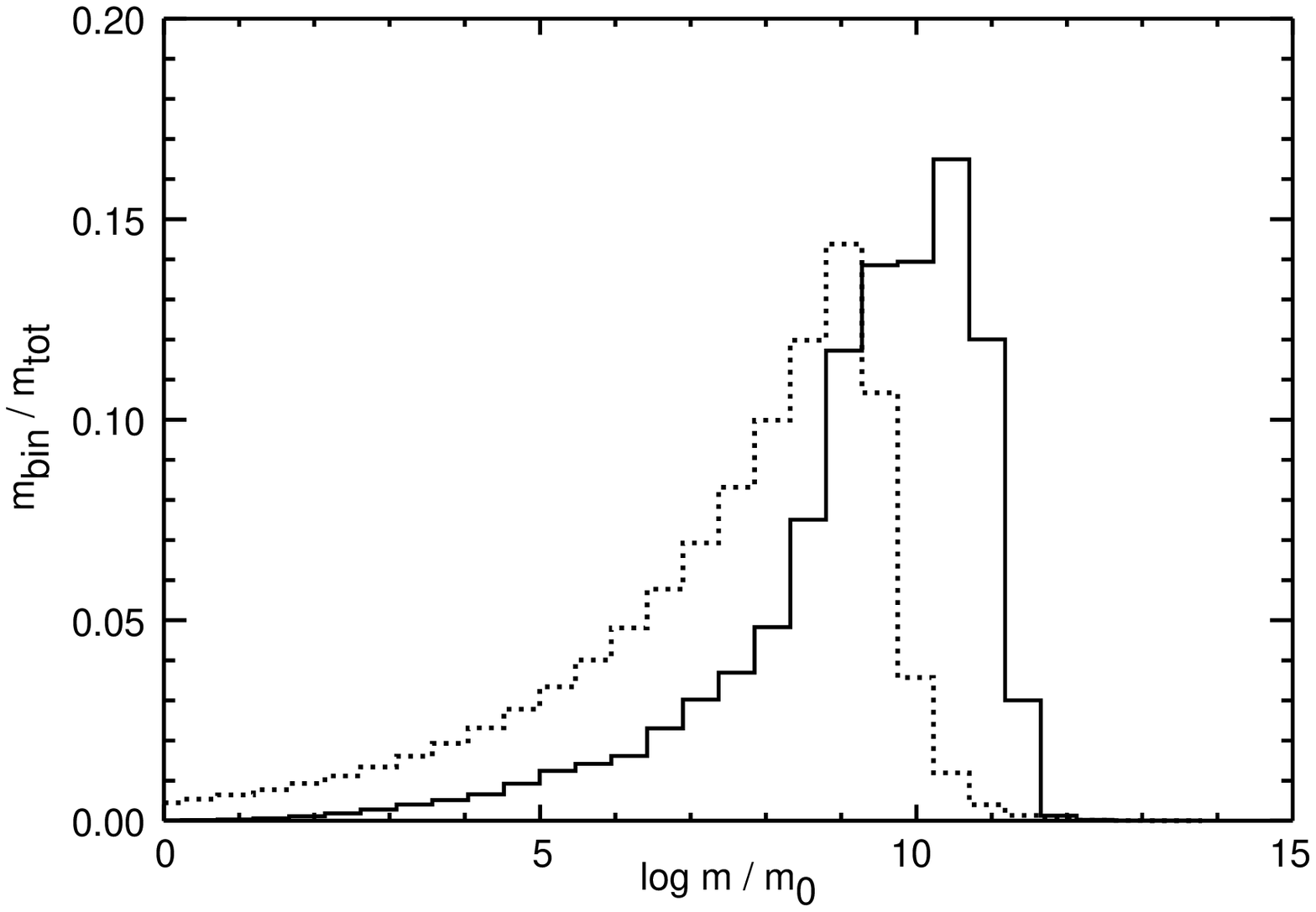}} 
\caption{Evolution of the total dust 
  mass distribution for the times shown in Fig. \ref{sPIspheres}. }
\label{dPIspheres}
\end{figure*}

\begin{figure*}
\hbox to \textwidth {\includegraphics*[width=80mm]{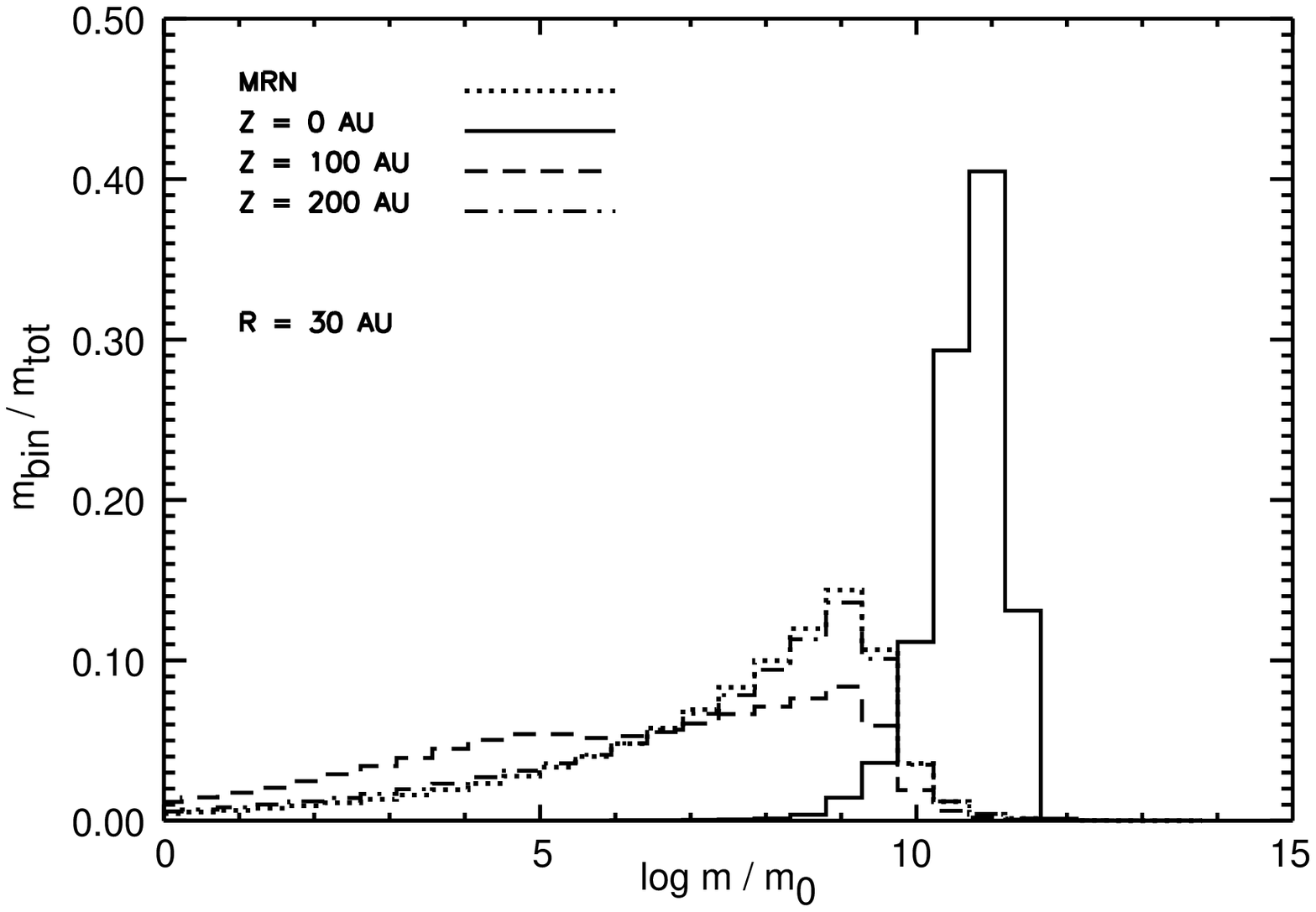}
             \hfil   \includegraphics*[width=80mm]{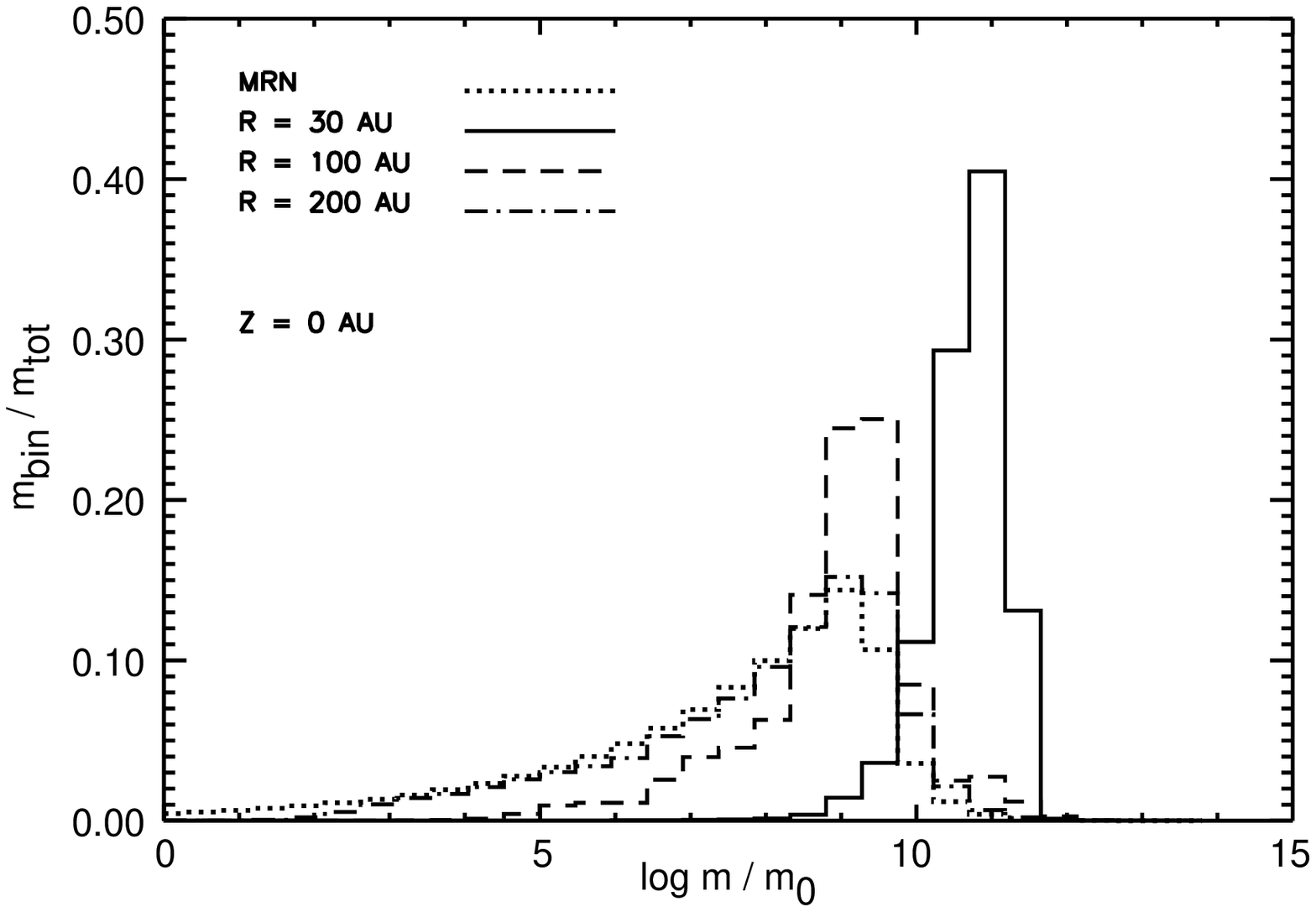}}
\caption{Model 3MS, 11400~yr. Left: Dust mass distribution at selected
  heights above the disk's midplane for $r=30$\,AU. Right: Dust mass
  distribution at selected radial distances for $z=0$\,AU.}
\label{PIseldist}
\end{figure*}

\begin{figure*}
\hbox to \textwidth {\includegraphics*[width=80mm]{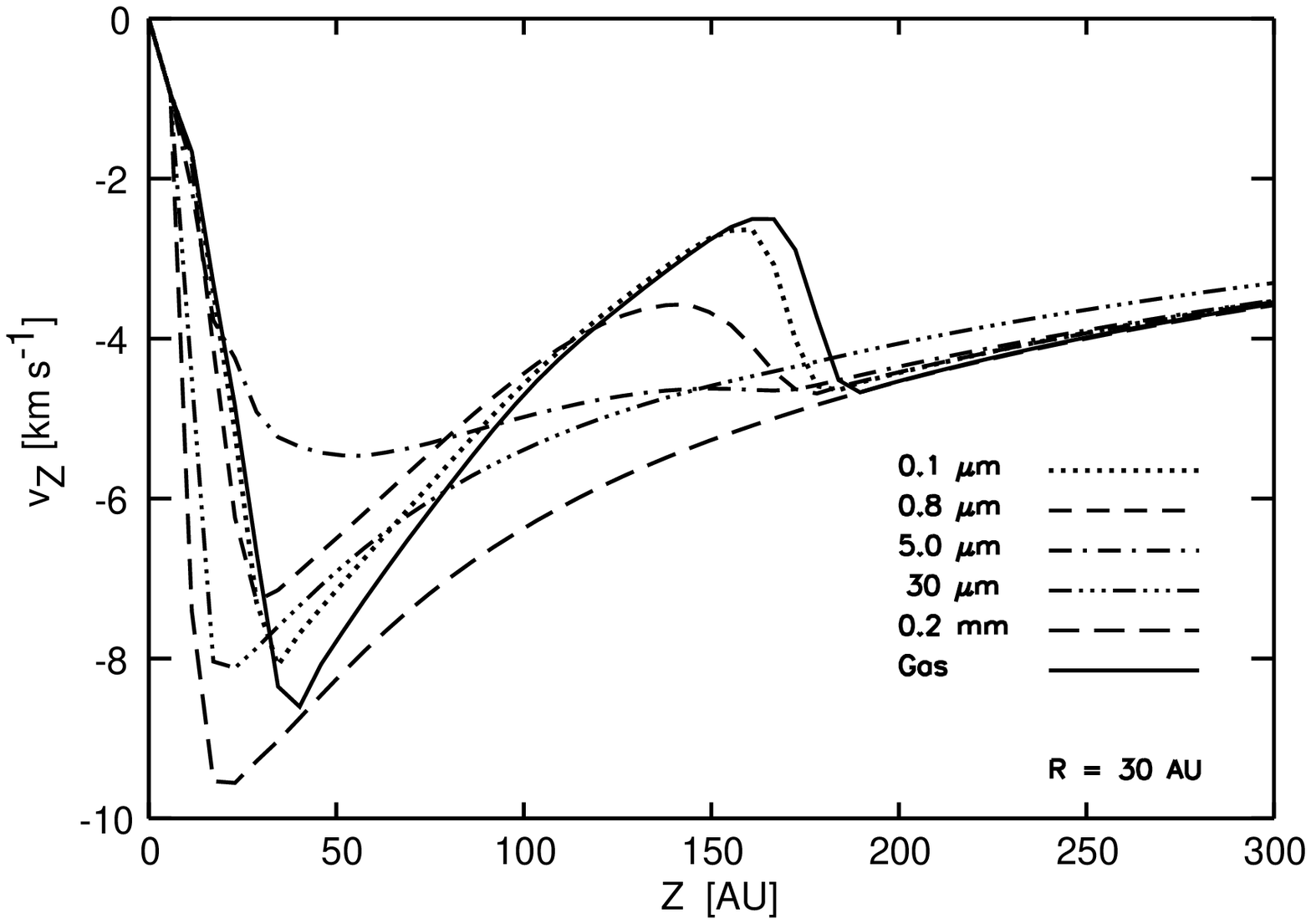}
             \hfil   \includegraphics*[width=80mm]{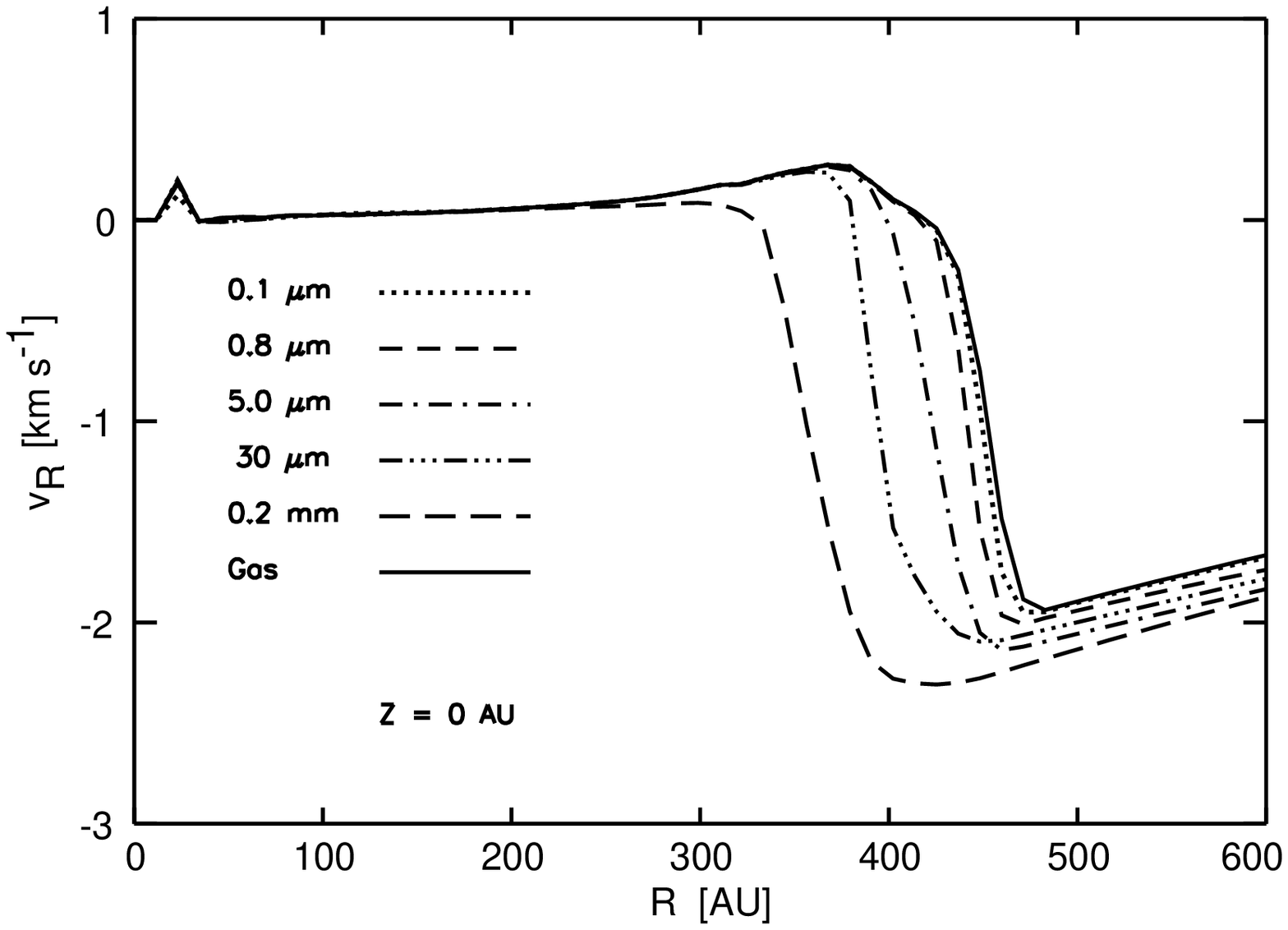}}
\caption{Model 3MS, 11400~yr. Left: Vertical components of velocity
  of selected dust components at $r=30$\,AU. Right: Radial velocities
  of selected dust components in the equatorial plane ($z=0$\,AU).}
\label{PIvshock}
\end{figure*}

The following numerical calculations were conducted with three nested grids of
increasing resolution of factor two each. The individual grids span
$60 \times 60$
zones. The dust size distribution is sampled by $N=30$ discrete dust species.
Table \ref{TAressimtab} summarizes selected results of the simulations.

\begin{table}[h]
\begin{center}
\caption{SELECTED RESULTS}
\hspace{0mm}\\
  \begin{tabular}{ccccc}
    \tableline\tableline\\[-3mm]
          & $M_*$       & $L_*$       &             &                   \\
    Model & [M$_\odot$] & [L$_\odot$] & $M_d / M_*$ & $t_s / t_{\rm ff}$\\[1mm]
    \tableline\\[-3mm]
      1MS & 0.77 &  3.7 & 0.30 & 2.5\\
      3MS &  2.2 &   87 & 0.36 & 2.3\\
      5MS &  3.7 &  194 & 0.35 & 2.1\\
     10MS &  8.2 & 1839 & 0.22 & 1.9\\
 1MS\_PCA & 0.77 &  4.0 & 0.30 & 2.5\\
 3MS\_PCA &  2.3 &   59 & 0.30 & 2.3\\
 5MS\_PCA &  3.7 &  270 & 0.35 & 2.2\\
10MS\_PCA &  7.9 & 1639 & 0.27 & 2.2\\
 1MS\_CCA & 0.77 &  2.9 & 0.30 & 3.1\\
 3MS\_CCA &  2.2 &   80 & 0.36 & 2.5\\
 5MS\_CCA &  3.8 &  188 & 0.32 & 2.3\\
10MS\_CCA &  8.0 & 1614 & 0.25 & 2.2\\[-6mm]
  \end{tabular}
\tablenotetext{}{{\sc Note.}---
  $M_*$ and $L_*$: Mass and luminosity of the central
  star; $M_d$: Mass of disk; $t_s$: Duration of simulation.}
\label{TAressimtab}
\end{center}
\end{table}

\subsection{Compact spherical dust grains}\label{SEsimcompact}

The first simulation applies the simple ``compact spherical grain'' dust model
(section \ref{SEcompact}). The collapse of the rotating molecular cloud core
is followed for about $10^4$~yr (about two initial free-fall times).
Figure \ref{sPIspheres} displays an evolutionary sequence of the gas
density and the gas velocity and Figure \ref{dPIspheres} the corresponding
total grain mass spectrum.

As evident in the lower right panel of Figure \ref{sPIspheres} two accretion
shock fronts have developed around the protostellar disk. The central mass
and core luminosity have attained values
2.2\,M$_\odot$ and 87\,L$_\odot$, respectively.
As shown in Figure \ref{dPIspheres}, coagulation removes the small dust
particles effectively from the size distribution. At the high mass end dust
grains of $\sim 50\,\mu$m are grown by coagulation. However, larger
particles do not form during the simulation. The integral size distribution
between 5\,nm and 5\,$\mu$m varies as $n(a) \propto a^{-3.1}$.

Figure \ref{PIseldist} (left panel) shows the dust mass spectrum at selected
heights above the protostellar accretion disk for a cylindrical
distance of 30\,AU.  At the disk midplane
(right panel) coagulation is very strong. Large grains are produced at the
cost of the small size end of the particle spectrum. At disk radii larger than
about 200 AU the effect of coagulation becomes less important. Only the small
grains are removed from the size spectrum by coagulation. However, in the
accretion shock just above the disk (Fig. \ref{PIseldist}, left panel) the
large grains are depleted relative to the low mass dust particles.

\begin{figure*}
\hbox to \textwidth {\includegraphics[width=80mm]{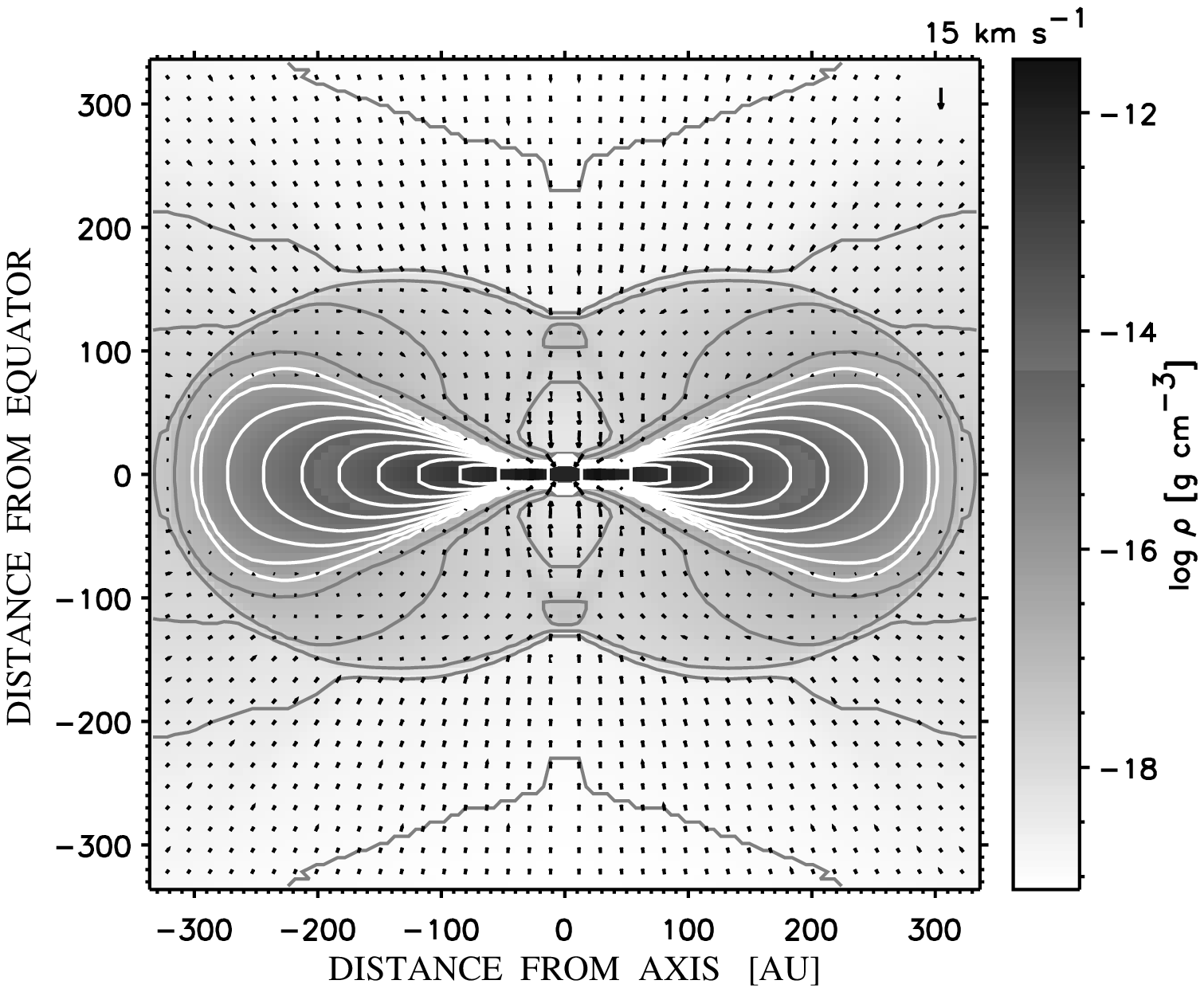}
            \hfil    \includegraphics[width=82mm]{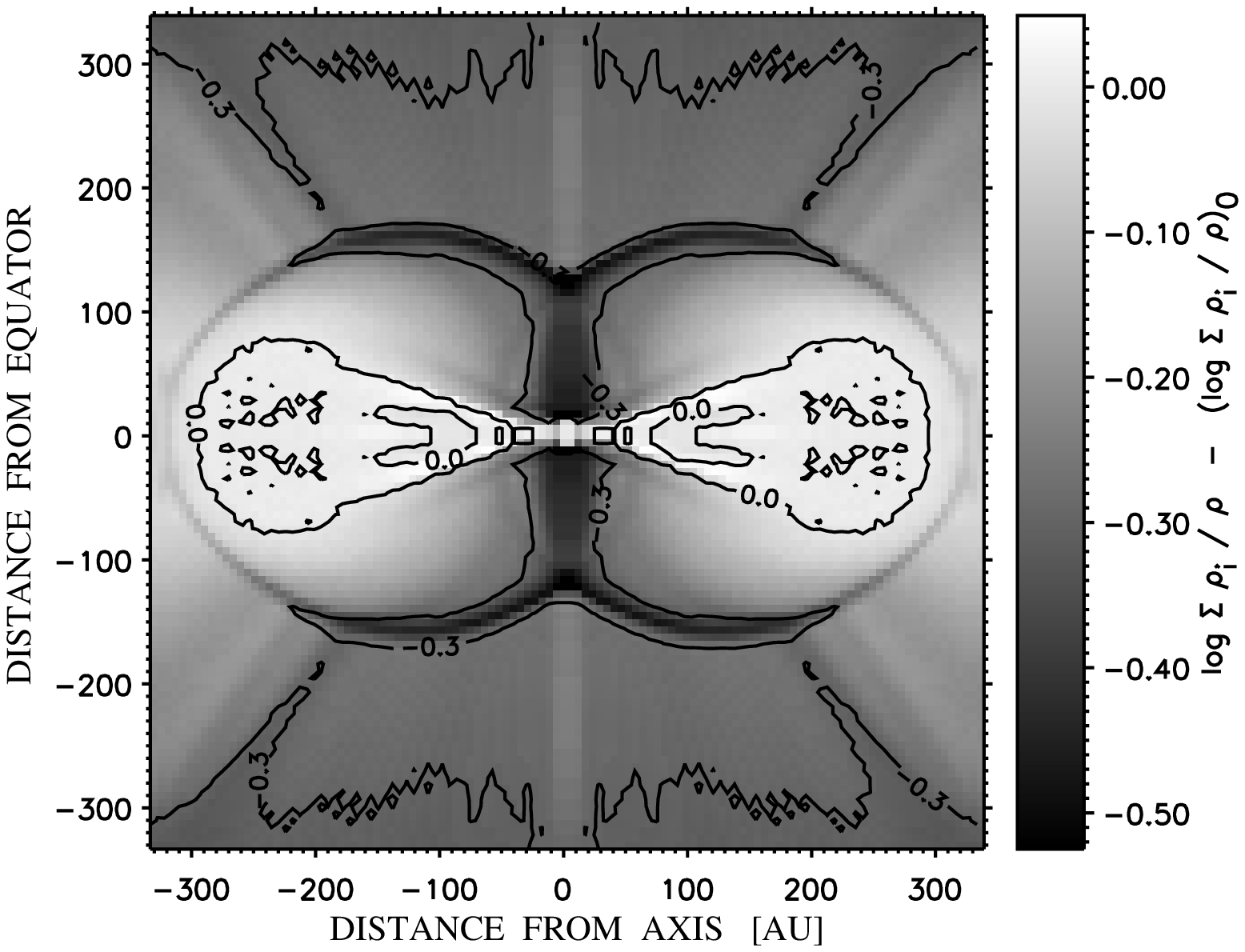}}
\caption{Model 3MS\_PCA, 11400~yr. Left: Mass density and velocity of the gas
  component.  Symbols and contour lines are as in Fig. \ref{sPIspheres}.
  Right: Variation of the dust to gas mass ratio in the accretion
  disk. \mbox{$(\sum\varrho_i / \varrho)_0$} is the initial value.
  The outer accretion shock (bunching of $\sim$2 contour lines
  in the left frame; surface of low dust to gas mass ratio in the
  right frame) at $z \approx \pm(125-150)$~AU is readily discernible.}
\label{PIpcafractals}
\end{figure*}

The velocities of selected dust grains through the accretion shock at
$r=30$\,AU (Fig. \ref{PIvshock}, left panel) show that dust grains of radii
$\approx 1\,\mu$m and above are coupled only loosely to the gas so that
significant relative velocities of several km\,s$^{-1}$ between these grains
and the smaller ones are achieved. This implies that coagulation is inhibited
and shattering might occur (the threshold velocity is 2.7\,km\,s$^{-1}$).
The densities are so low, however, that the shattering time scale is
$\sim 10^4$~yr. Thus, the depletion of the large grains must be
attributed to the fact that they pass quickly through this zone,
whereas the smaller grains slow down significantly.  This size dependent
gas--grain drift lowers
the dust to gas mass ratio by more than a factor of 2 between the two accretion
shocks (shown for the BPCA fractals, see Fig. \ref{PIpcafractals}, right
panel).

Note the differing velocities above the outer accretion shock ($z\approx
170$\,AU), caused by the size dependence of the grain opacity and thus
by differential radiative acceleration.  Without radiative acceleration the
grains should fall towards the equatorial plane faster than the gas,
because they are not pressure supported. Outside the outer accretion
shock the grains' mass distribution is still well approximated by the
initial MRN mass distribution
(Fig. \ref{PIseldist}, left panel). In outer regions of the cloud at
the equatorial plane the infalling material is shielded from the central
star by the disk. Hence, the grains
do indeed accrete with higher velocities than the gas component (Fig.
\ref{PIvshock}, right panel).

To quantify the effect of a modified dust size spectrum on the optical
properties of the protostellar matter in the accretion disk, Figure
\ref{PIevext} (left panel) compares the net specific extinction coefficient
for several locations in the accretion disk. Whereas the depletion of the
large grains behind the accretion shock does introduce minor modifications to
the extinction coefficient, coagulation at the disk's midplane causes an opacity
reduction of more than an order of magnitude for the near infrared to UV
extinction. From $\lambda=0.1$\,cm to 0.1\,mm the extinction coefficient
is increased by approximately the same amount. This behaviour is indicative
of the migration of the peak of the grain mass distribution to
higher masses due to coagulation.

To drive the coagulation beyond the initial upper grain size limit of
$5\,\mu$m, systematic relative velocities are needed, such as 
those which result from differential radiative acceleration,
relaxation behind the accretion shock and gravitative sedimentation.
Figure \ref{PIevext} (right panel) displays the resulting dust mass
distribution, when these contributions to the relative motions are
neglected. Obviously, turbulence and Brownian motion are sufficient to remove
the small grains, but they are not able to build up $\mu$m-sized particles
quickly.
Whereas for cloud clump masses of 1\,M$_\odot$ the differential
radiative acceleration of grains is negligible, it eventually becomes
the most important mechanism for creating velocity differences
between grains outside the accretion shock fronts when
larger clump masses are considered.

\begin{figure*}
\hbox to \textwidth {\includegraphics[width=80mm]{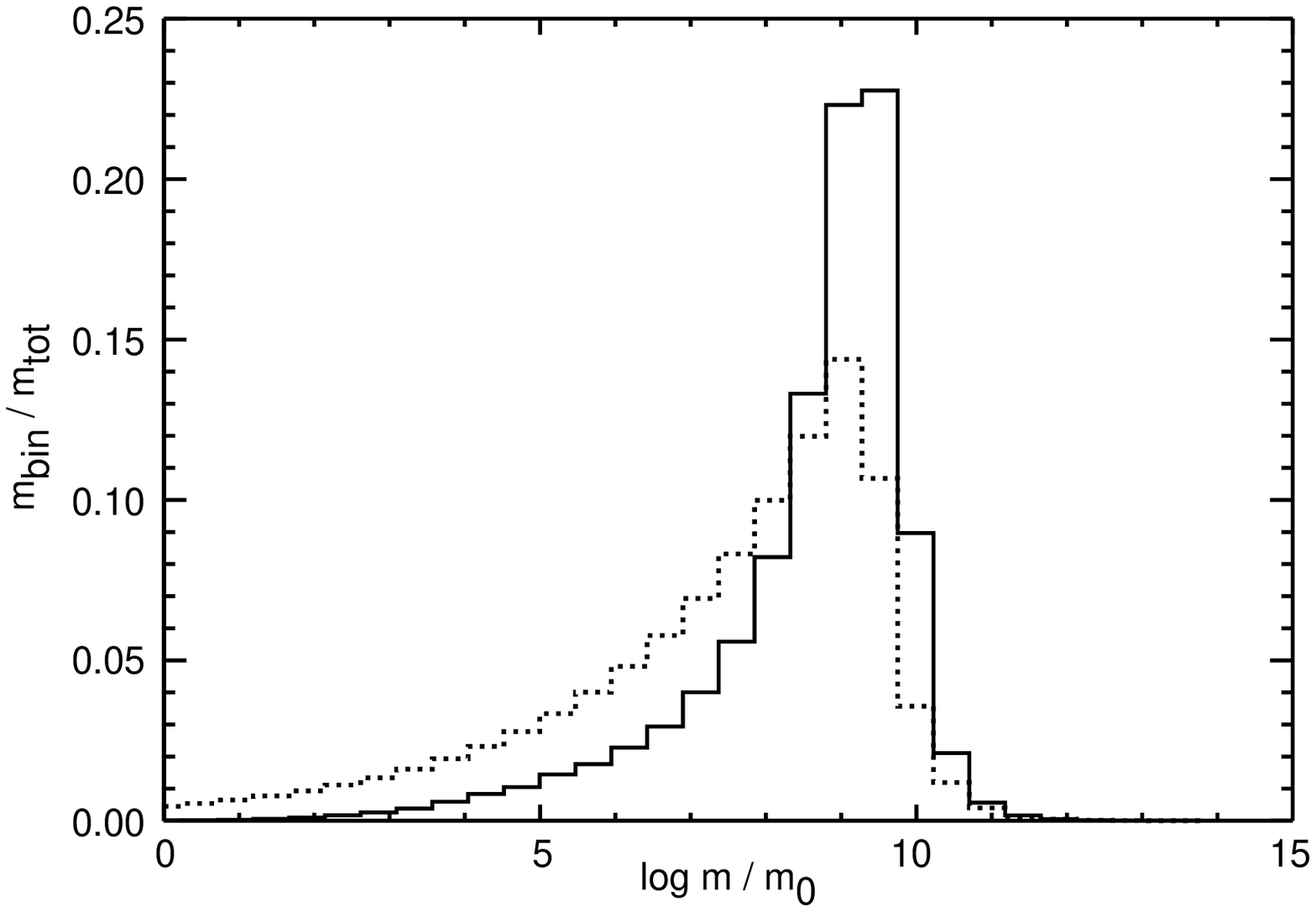}
            \hfil    \includegraphics[width=80mm]{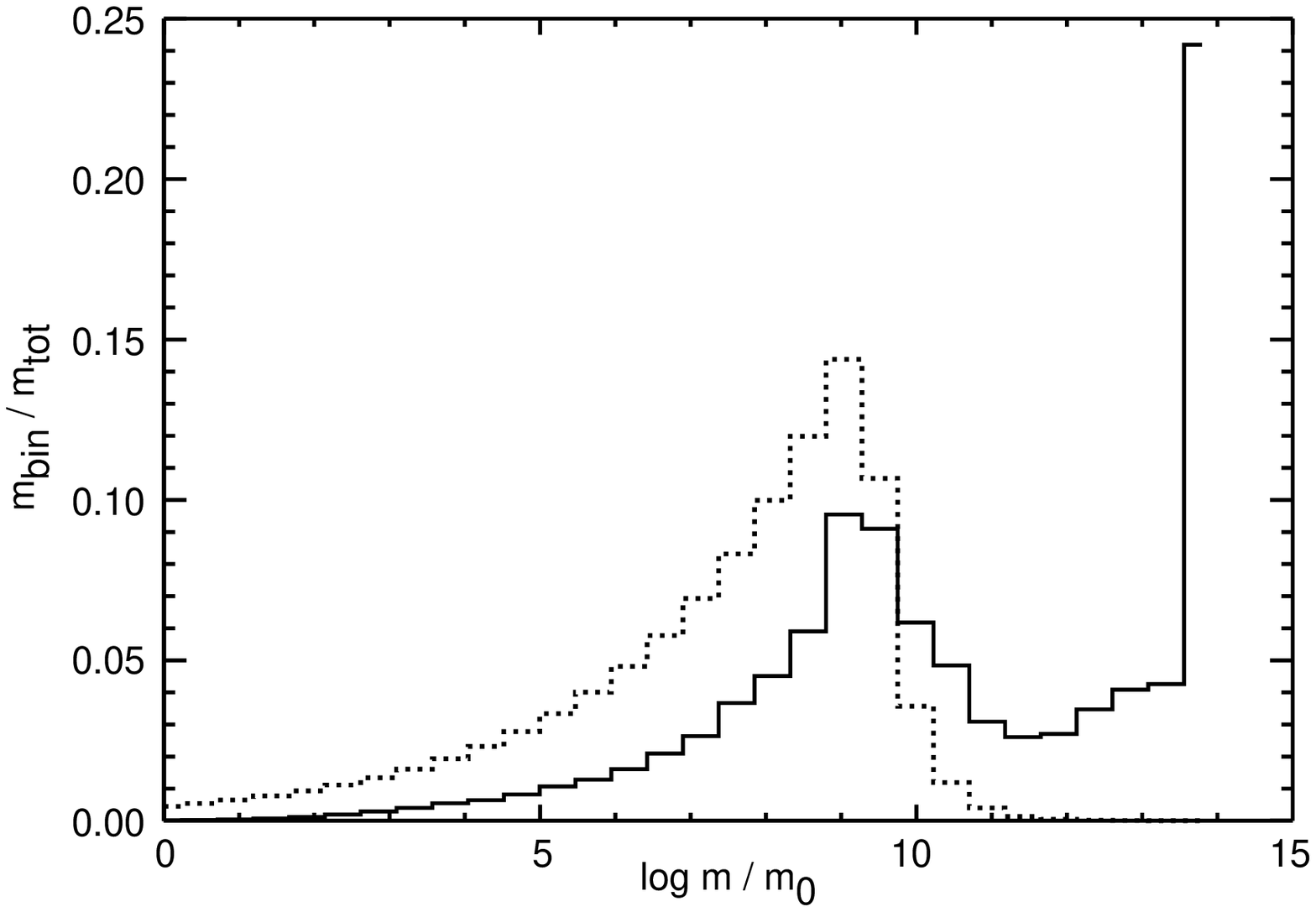}}
\caption{Overall mass spectrum of the grains ({\it solid lines})
  compared to the initial MRN mass distribution ({\it dotted lines}).
  Left: The critical sticking velocity is reduced by a
  factor of 10 compared to model 3MS.
  Right: For comparison, the critical sticking velocity is set to infinity.}
\label{PImiestick}
\end{figure*}

\begin{figure*}
\hbox to \textwidth {\includegraphics[width=78mm]{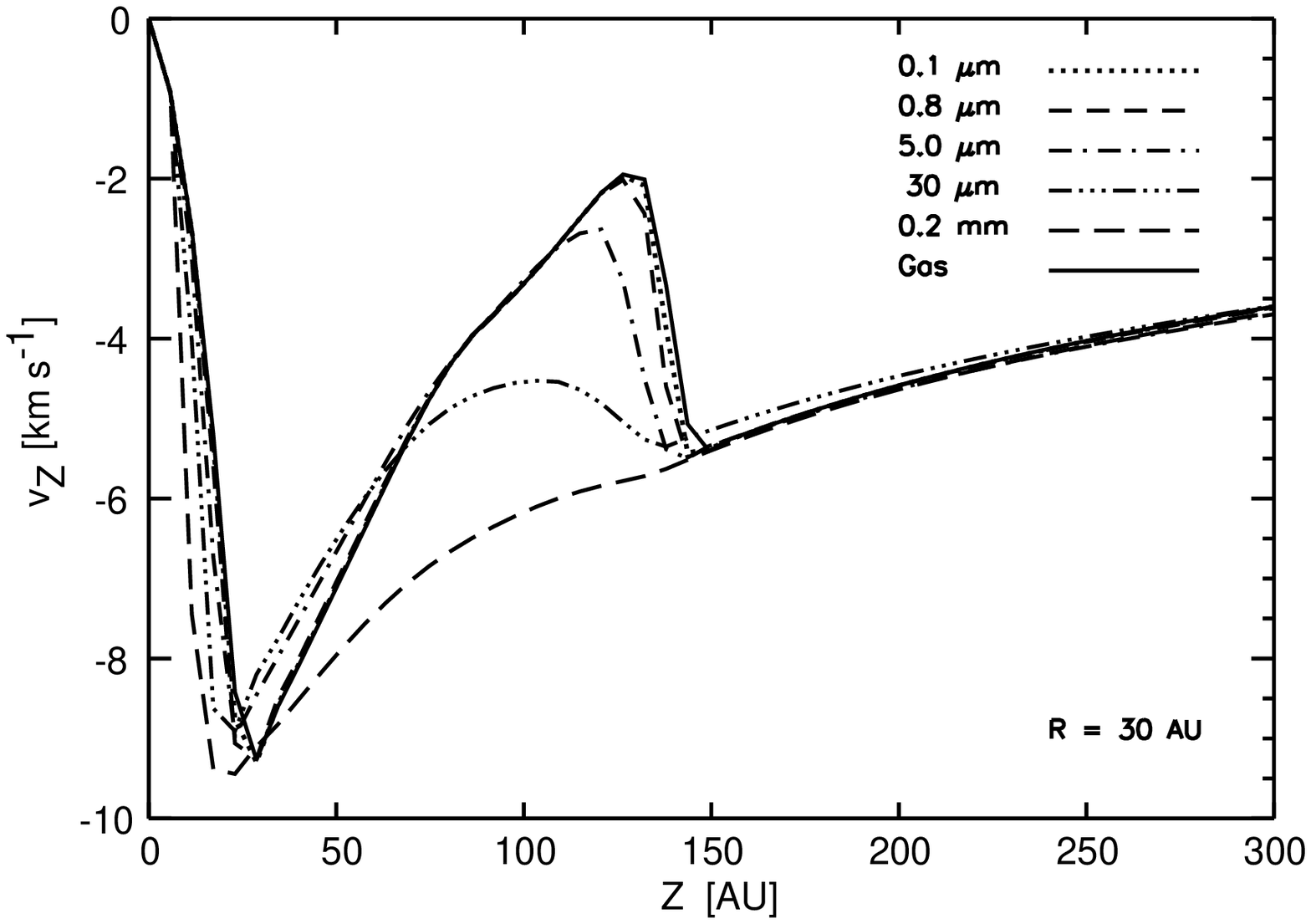}
            \hfil    \includegraphics[width=82mm]{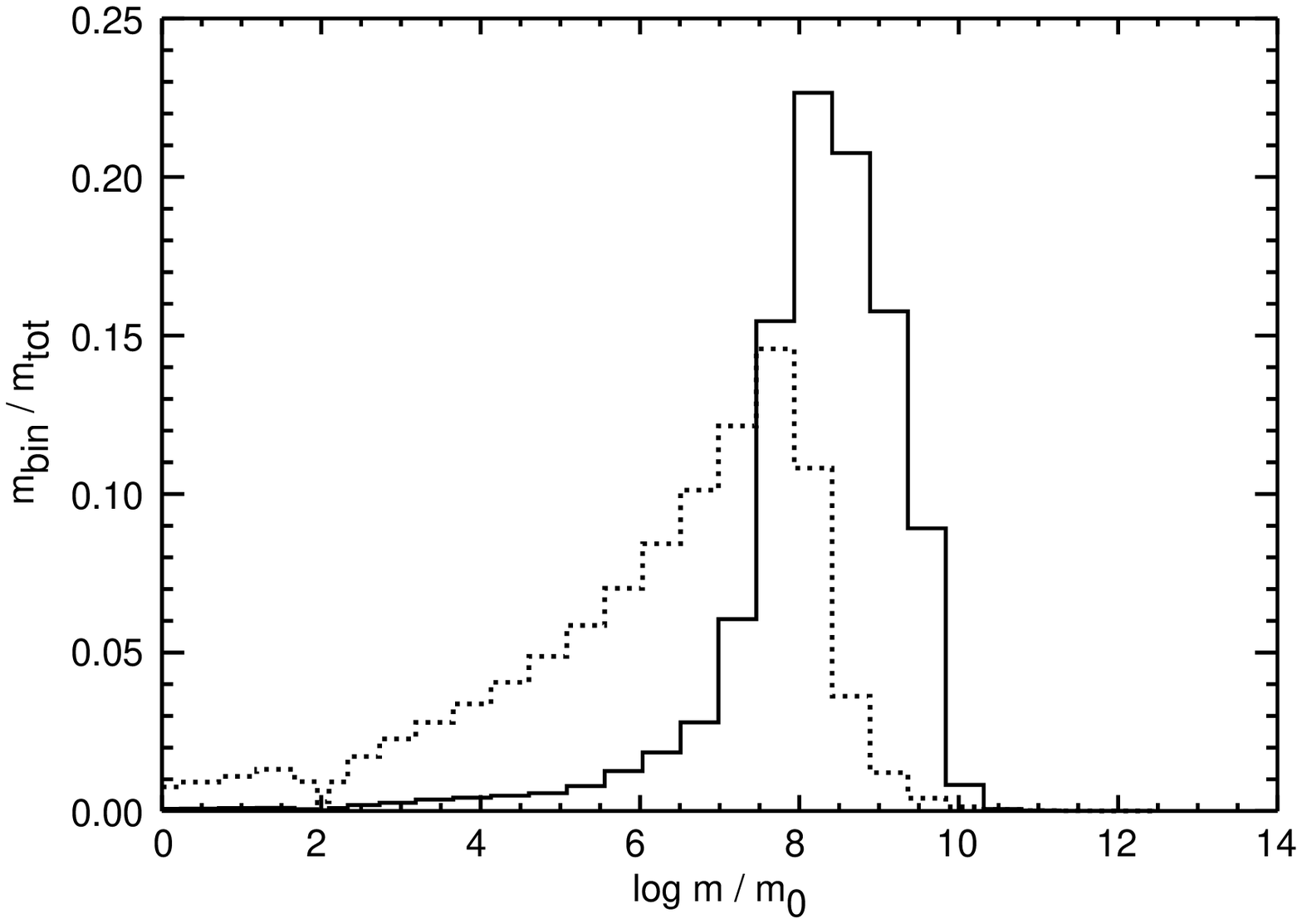}}
\caption{Model 3MS\_PCA, 11400~yr. Left: Vertical components of velocity
  of selected dust components at $r=30$\,AU. Right: Overall
  mass spectrum of the coagulated dust grains ({\it solid line}) compared to
  the initial MRN mass distribution ({\it dotted line}).}
\label{PIpcavel}
\end{figure*}

\begin{figure*}
\hbox to \textwidth {\includegraphics*[width=77mm]{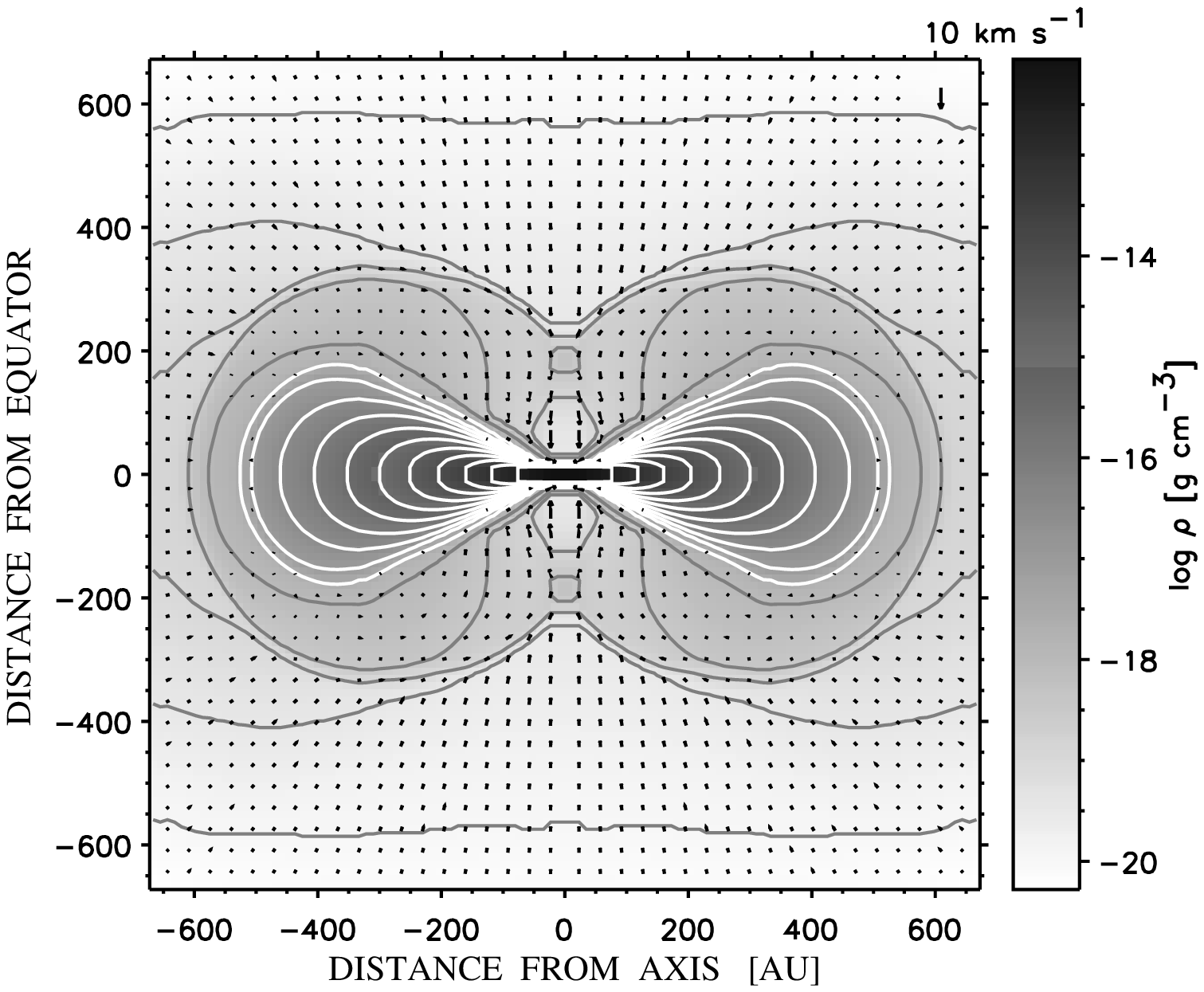}
            \hfil    \includegraphics*[width=83mm]{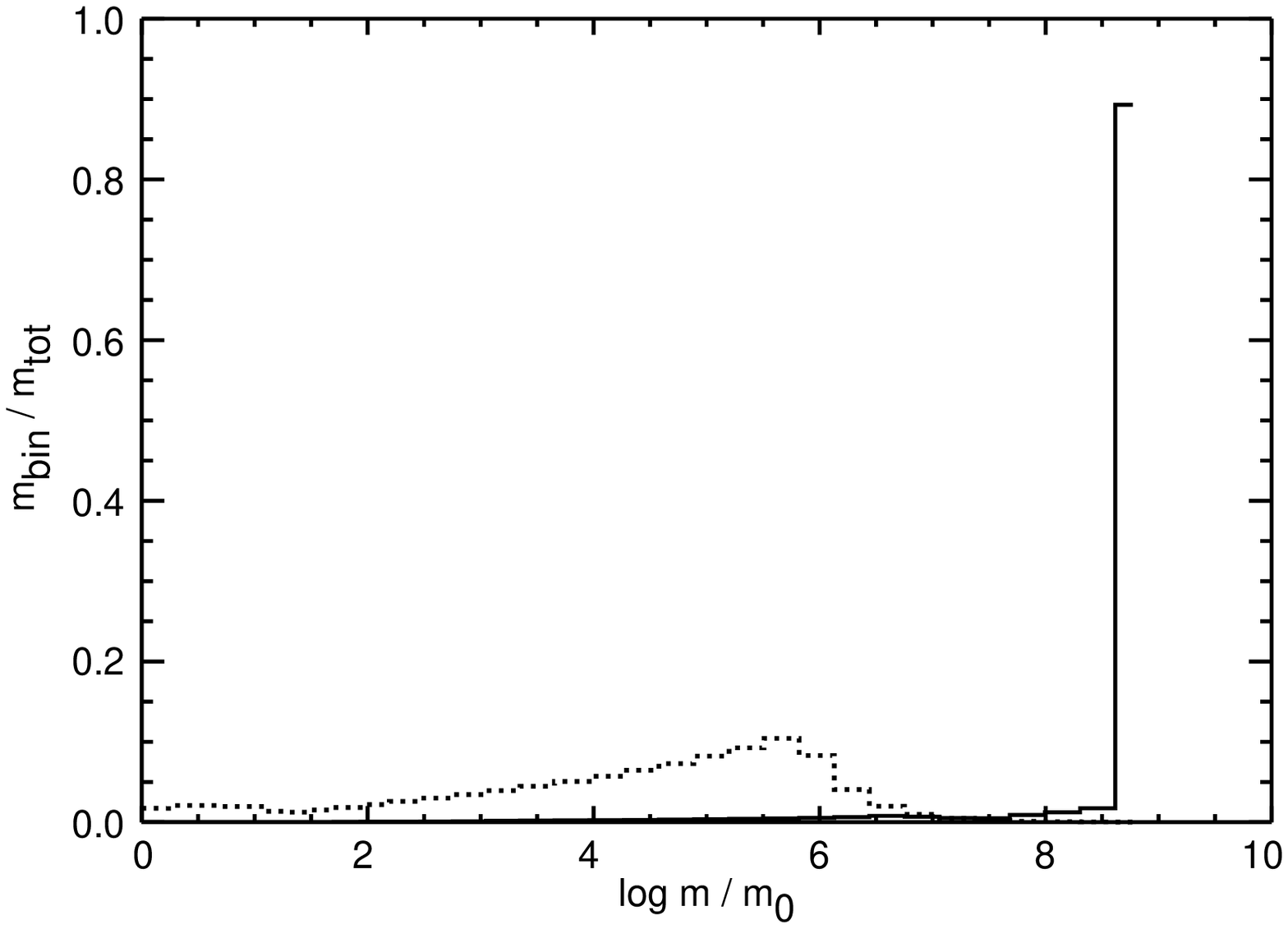}}
\caption{Model 3MS\_CCA, 12600~yr. Left: Gas density and velocity structure.
  Right: Overall mass spectrum of the coagulated dust grains ({\it solid line})
  compared to the initial MRN mass distribution ({\it dotted line}).}
\label{PIccafractals}
\end{figure*}

\begin{figure*}
\hbox to \textwidth {\includegraphics*[width=80mm]{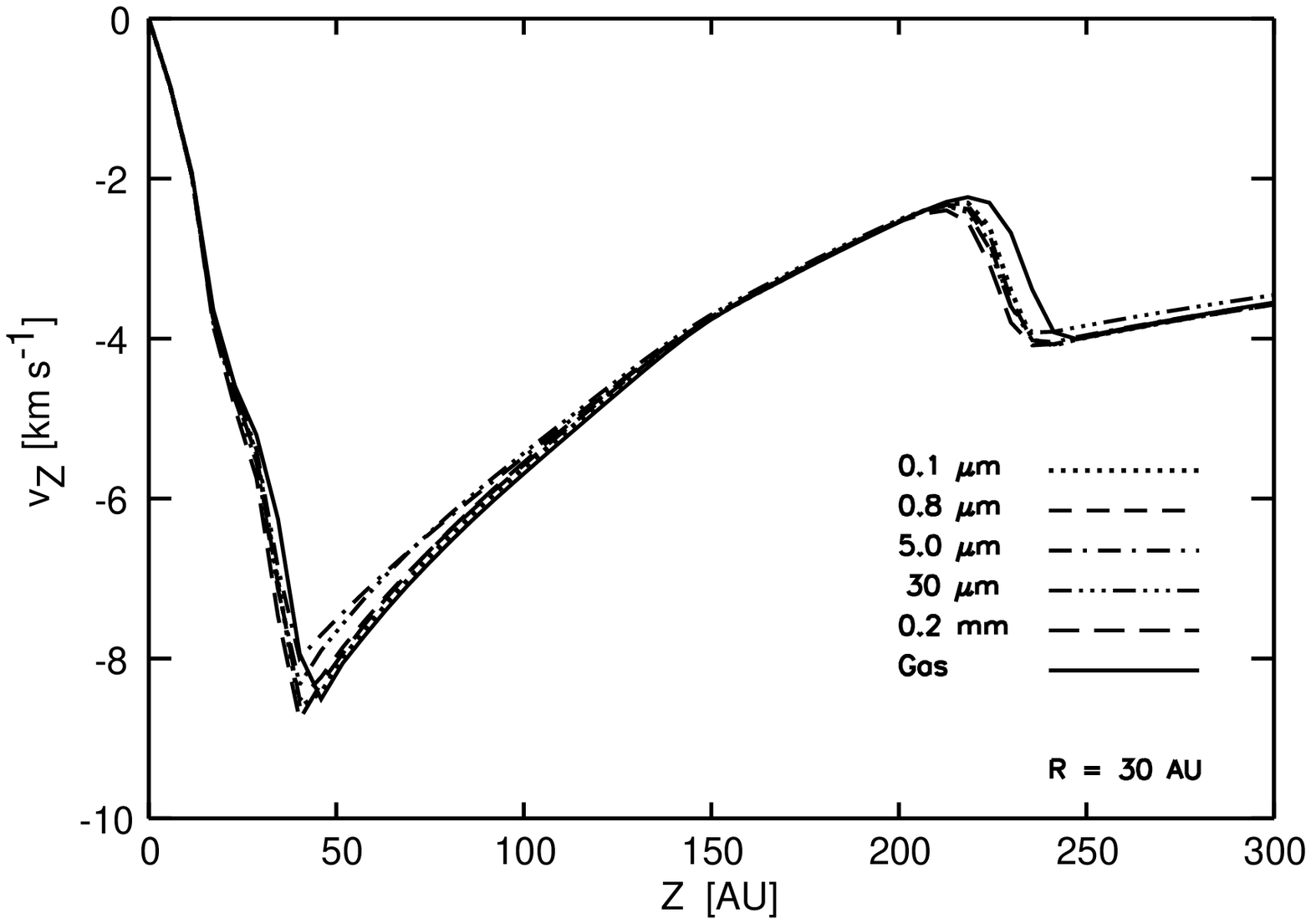}
            \hfil    \includegraphics*[width=80mm]{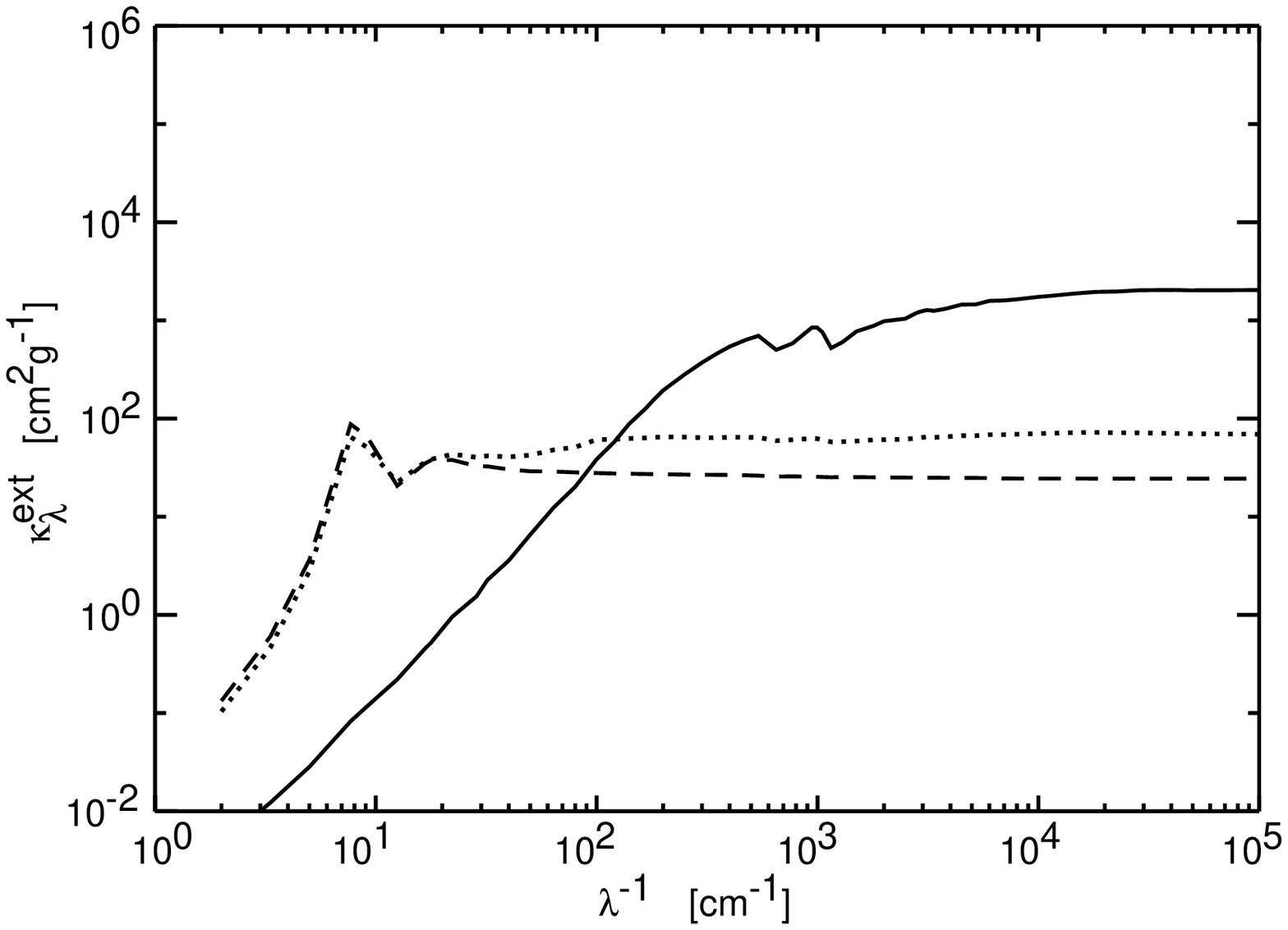}}
\caption{Model 3MS\_CCA, 12600~yr. Left: Vertical components of velocity
  of selected dust particles at $r=30$\,AU.
  Right: The net specific extinction coefficient
  of unmodified dust ({\it solid line}) compared to the numerical results
  for the equatorial midplane
  at $r=300$\,AU ({\it dotted line}) and $r=30$\,AU ({\it dashed line}).}
\label{PIccavel}
\end{figure*}

\begin{figure*}
\hbox to \textwidth {\includegraphics*[width=77mm]{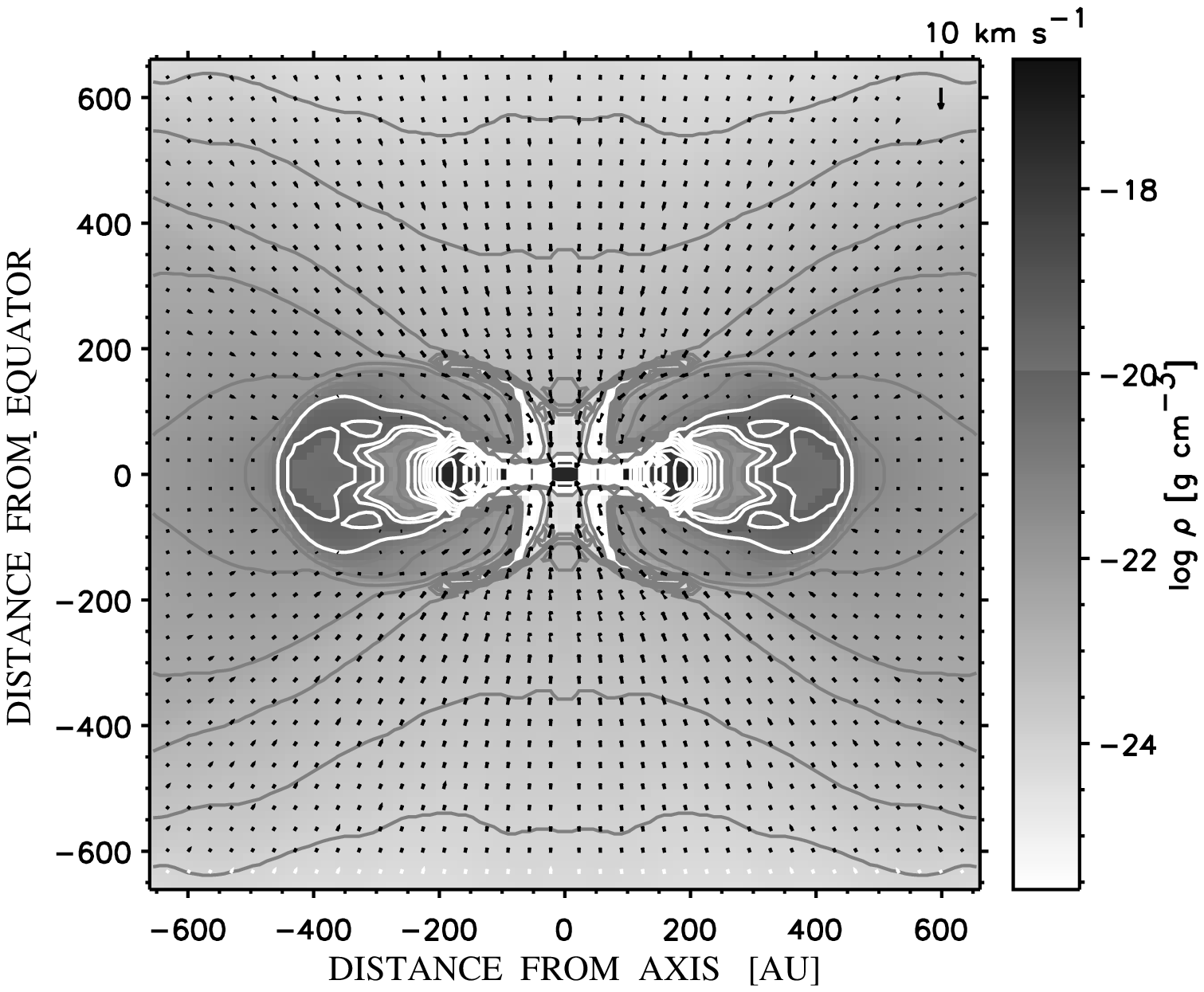}
            \hfil    \includegraphics*[width=83mm]{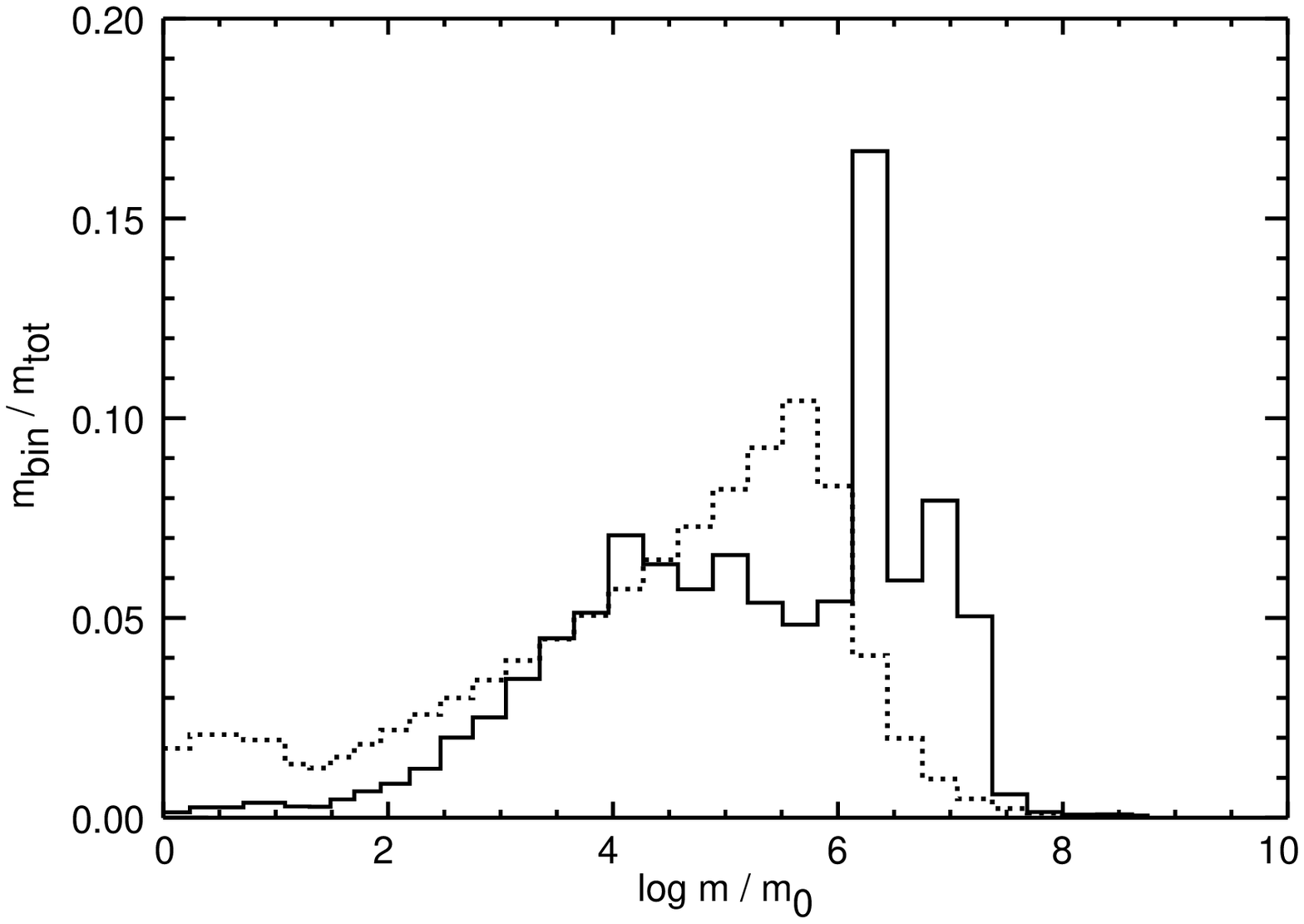}}
\caption{Model 3MS\_CCA\_S10, 12700~yr. Left: Mass density and velocity of
  dust particles with $r_{ex}=5\,\mu$m.
  Right: Overall mass spectrum of the coagulated dust grains ({\it solid line})
  compared to the initial MRN mass distribution ({\it dotted line}).}
\label{PIccastick}
\end{figure*}

\subsection{Influence of the sticking properties of the grains}
\label{SEinflstick}

How is coagulation affected by the sticking strength of the dust
particles? As pointed out in the introduction, this material property is not
yet well understood. Reports on experimental studies indicate larger values for
the critical sticking velocity than theoretical models predict (Poppe \& Blum
1997). Ices on the grain surfaces play an important role (Supulver et al.
1997). To investigate this effect we conducted comparison simulations with 
different
sticking strengths. First, the critical sticking velocities were reduced to
the conservative theoretical values (without the factor 10 correction
for the experimental results, see section \ref{SEspstick}).

Figure \ref{PImiestick} (left panel) displays the total mass distribution.
Obviously, grain growth is reduced; the maximum grain mass is smaller
by a factor
of about 10. When the critical sticking velocity is set to infinity,
i.e. the grains stick at every encounter, grains up to the actual bin limit
with radii of about 0.2\,mm are grown. This demonstrates that the material
parameters play an extremely important role in defining the upper mass limit up
to which the dust grains are able to grow during the formation of an accretion
disk.

\subsection{Fractal BPCA particles}

Our next approximation treats the dust grains as fractal coagulated particles
(BPCA particles). The same initial and boundary conditions as in the previous
calculation are used. At the end of the simulation the mass of the central
object is 2.3\,M$_\odot$ with a luminosity of 59\,L$_\odot$. The overall
evolution is qualitatively similar to the calculation with compact dust
grains (Fig. \ref{sPIspheres} and Fig. \ref{dPIspheres}).
Figure \ref{PIpcafractals} shows the mass density and
velocity of the gas component (left panel) and the dust to gas mass ratio
(right panel). Again, an accretion shock has developed. In this accretion shock
the dust to gas mass ratio is reduced by a factor of about 3 compared to the
initial value due to size dependent advection.

The specific cross section of large BPCA coagulates is about a factor of 5
larger than for compact spheres of the same mass (section \ref{SEfrac}).
Only particles with radii of about 10\,$\mu$m and larger decouple from the
motion of the gas flow in the accretion shock (Fig. \ref{PIpcavel}, left
panel; compare to Fig. \ref{PIvshock}). Coagulation in the equatorial
plane proceeds at a rate similar to the simple compact spherical dust
model (Fig.
\ref{PIpcavel}, right panel). As in the case of spherical grains no dust
particles larger than several 10\,$\mu$m are grown by coagulation. This can be
attributed again to a finite critical sticking velocity (see section
\ref{SEinflstick}).

\subsection{Fractal BCCA particles}

Finally, BCCA grains are the most fluffy particles dealt with in these
simulations. Figure \ref{PIccafractals} (left panel) shows the density and the
velocity of the gas component in Model 3MS\_CCA. The overall distribution is
similar to the previously discussed models (3MS and 3MS\_PCA). However, grain
coagulation is enormously strong. Grains as large as 0.2\,mm (at the limiting
end of the binning) are grown (Fig. \ref{PIccafractals}, right panel).
Almost all the grain mass resides in the most massive bin. Because of the
fluffy structure of the BCCA grains the dust remains well coupled to the gas,
even in the accretion shock (Fig. \ref{PIccavel}, left panel). Thus, relative
velocities between the dust particles remain very low.

The optical properties of the gas--dust mixture at several locations in the
accretion disk are plotted in Figure \ref{PIccavel} (right panel). In the
equatorial plane the net specific extinction coefficient
$\kappa^{\rm ext}_\lambda$ is
lowered by more than an order of magnitude from the near infrared to UV
wavelengths. From 1\,mm to 100\,$\mu$m the extinction is enhanced. In the
accretion shock only minor modifications can be ascertained.

A theoretical shattering model differing somewhat from the one
discussed above has been developed by Dominik \& Tielens (1997).
In order to test its effect on our simulations we used this model to
compute the evolution of BCCA grains. According to Dominik \&
Tielens (1997) the shattering velocity
is proportional to the critical sticking velocity. It also depends
on the number of contact points between the two colliding dust grains.
In our ignorance of this quantity we assume that two dust grains
always have 10 contact points. Figure \ref{PIccastick} (left panel) displays
the density and velocity of dust grains with reduced radii $r_{ex}=5\,\mu$m.
In the accretion shock near to the axis of rotation these dust particles are
destroyed by shattering encounters. As can be seen in the total dust mass
distribution (Fig. \ref{PIccastick}, right panel),
the largest dust grains with radii of 0.2\,mm
are not formed. We attribute this to frequent shattering encounters;
the sticking properties are identical to those used in model 3MS\_CCA.

\subsection{Synthetic emission maps and spectra}

\begin{figure*}
\hbox to \textwidth {\includegraphics*[width=80mm]{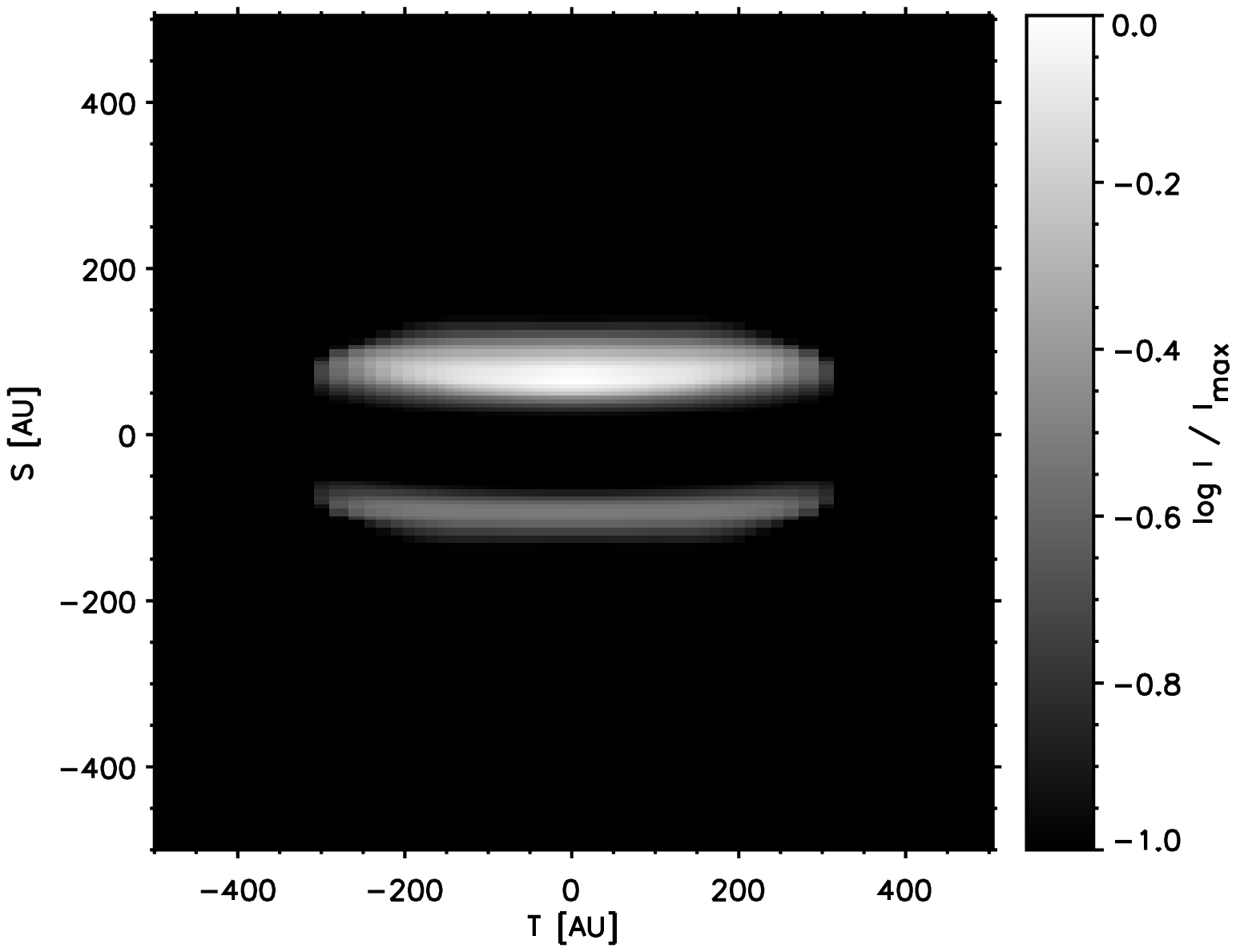}
         \hfil       \includegraphics*[width=80mm]{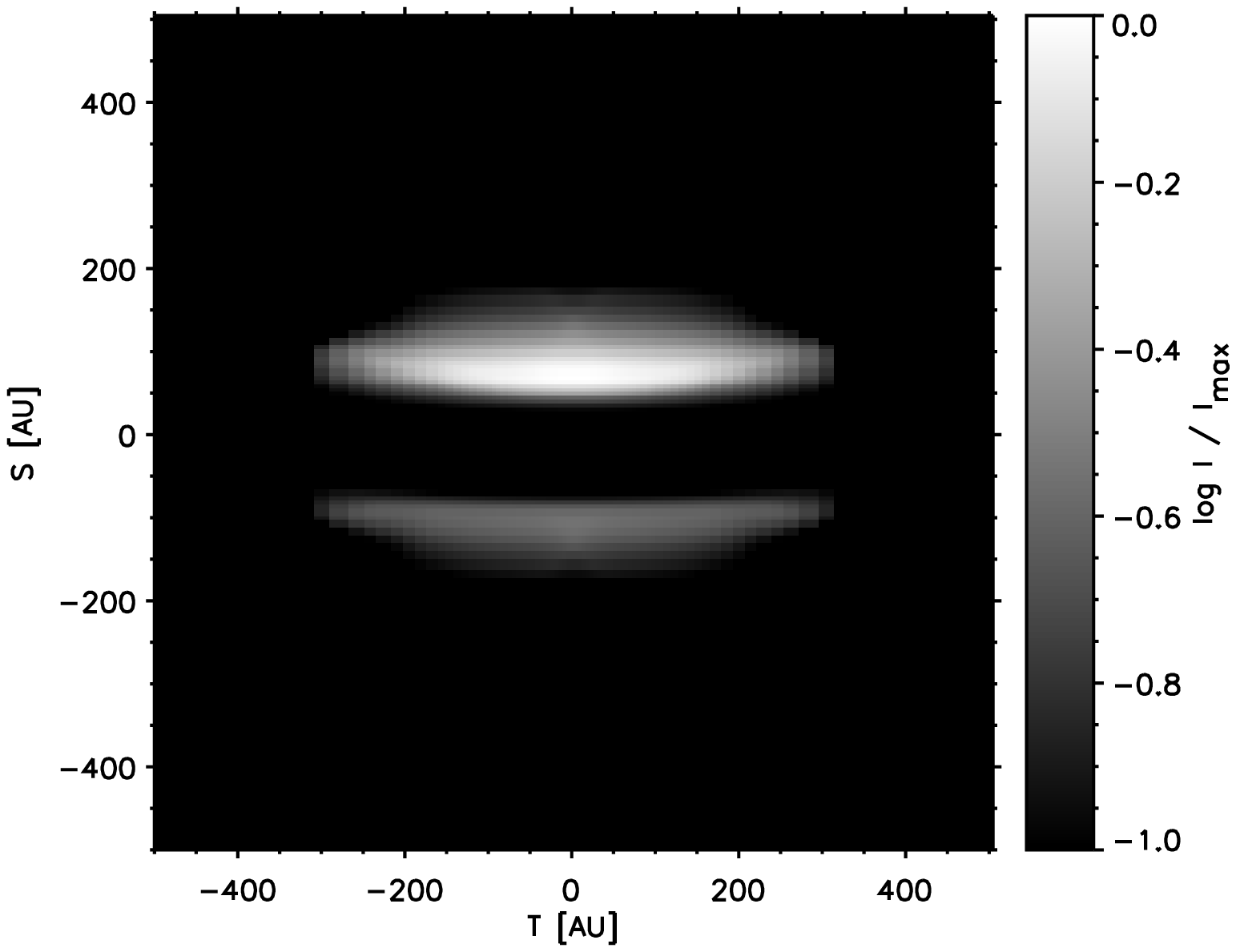}}
\caption{Dust continuum emission of model 3MS, 11400~yr at
  an angle of inclination of $\theta=85^\circ$.
  Left: Continuum emission map at $\lambda=3.6\,\mu$m using
  coagulated dust as calculated.
  Right: Comparison emission map using unmodified dust.}
\label{PIemsphere}
\end{figure*}

\begin{figure*}
\hbox to \textwidth {\includegraphics*[width=80mm]{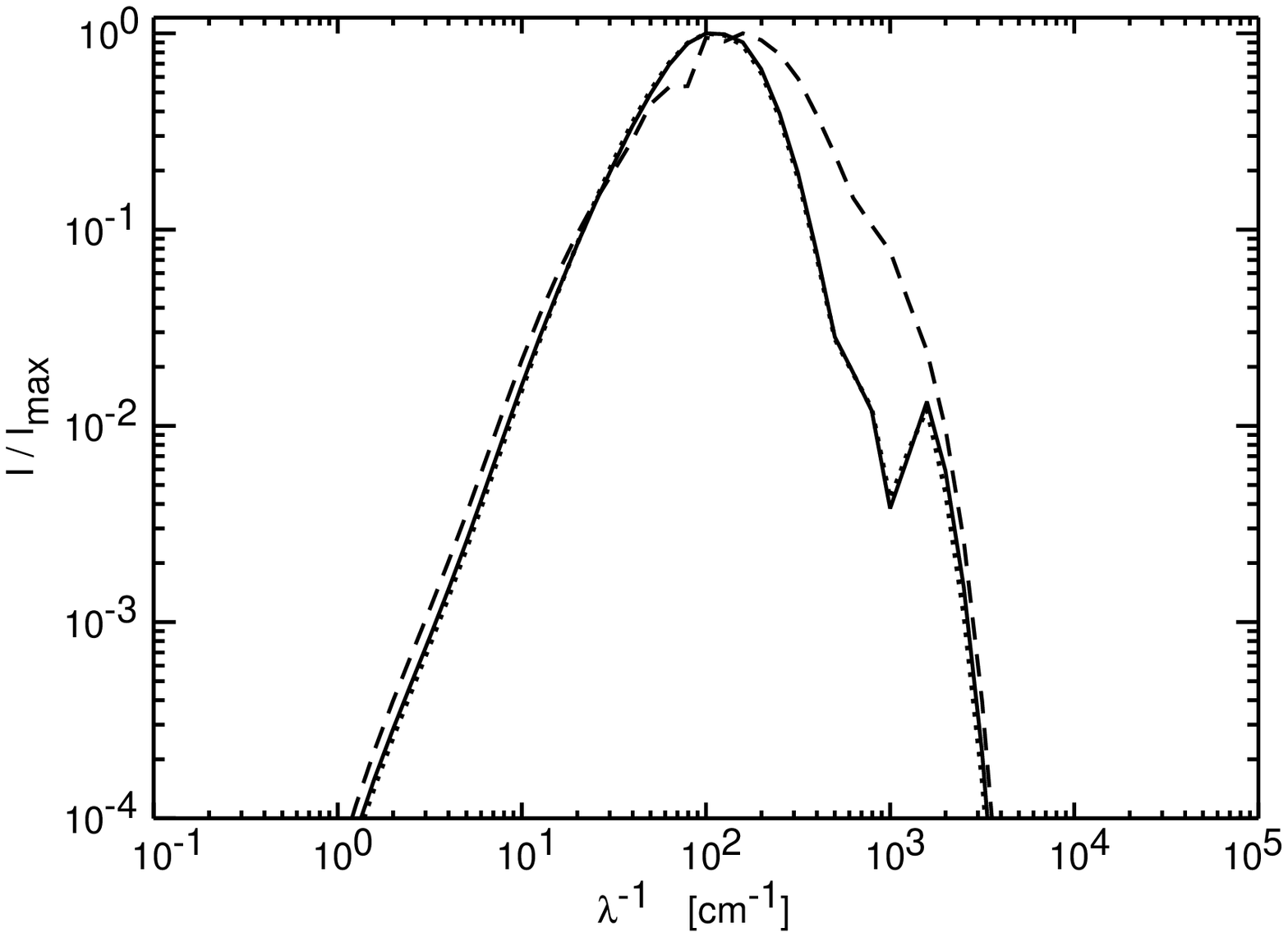}
         \hfil       \includegraphics*[width=80mm]{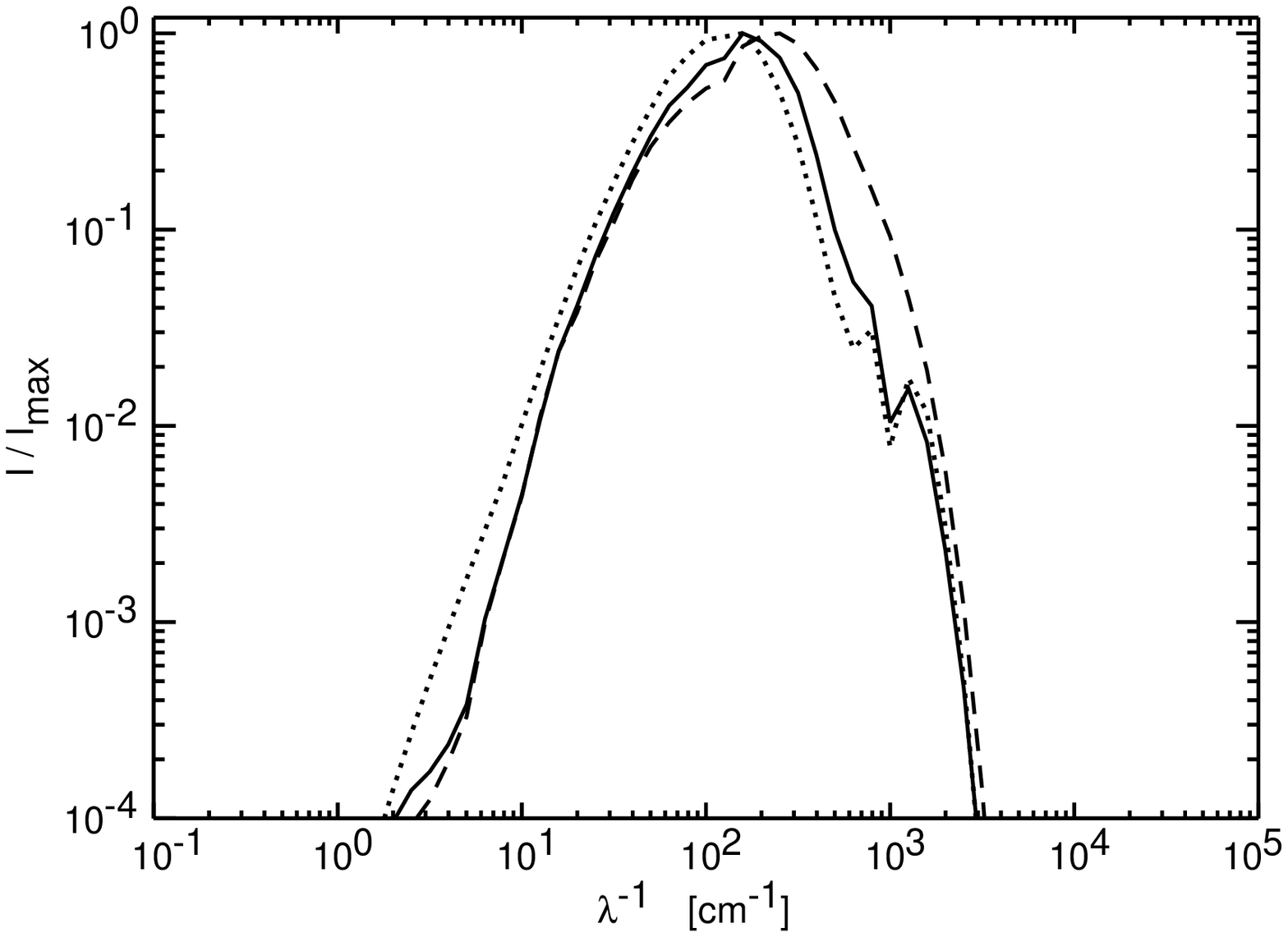}}
\caption{Emission spectra resulting from simulations leading to
  coagulated dust ({\it solid line}), with unmodified
  dust ({\it dotted line}), or assuming monodisperse dust
  ({\it dashed line})
  with $a=22\,\mu$m (model 3MS) and $r_{ex}=0.2\,$mm (3MS\_CCA).
  The rotation axis of the disk is inclined at an angle
  $\theta=85^\circ$ with respect to the line of sight.
  Left: Model 3MS, 11400~yr.  Right: Model 3MS\_CCA, 12600~yr.}
\label{PIemsphereccaspec}
\end{figure*}

\begin{figure*}[ht]
\hbox to \textwidth {\includegraphics*[width=80mm]{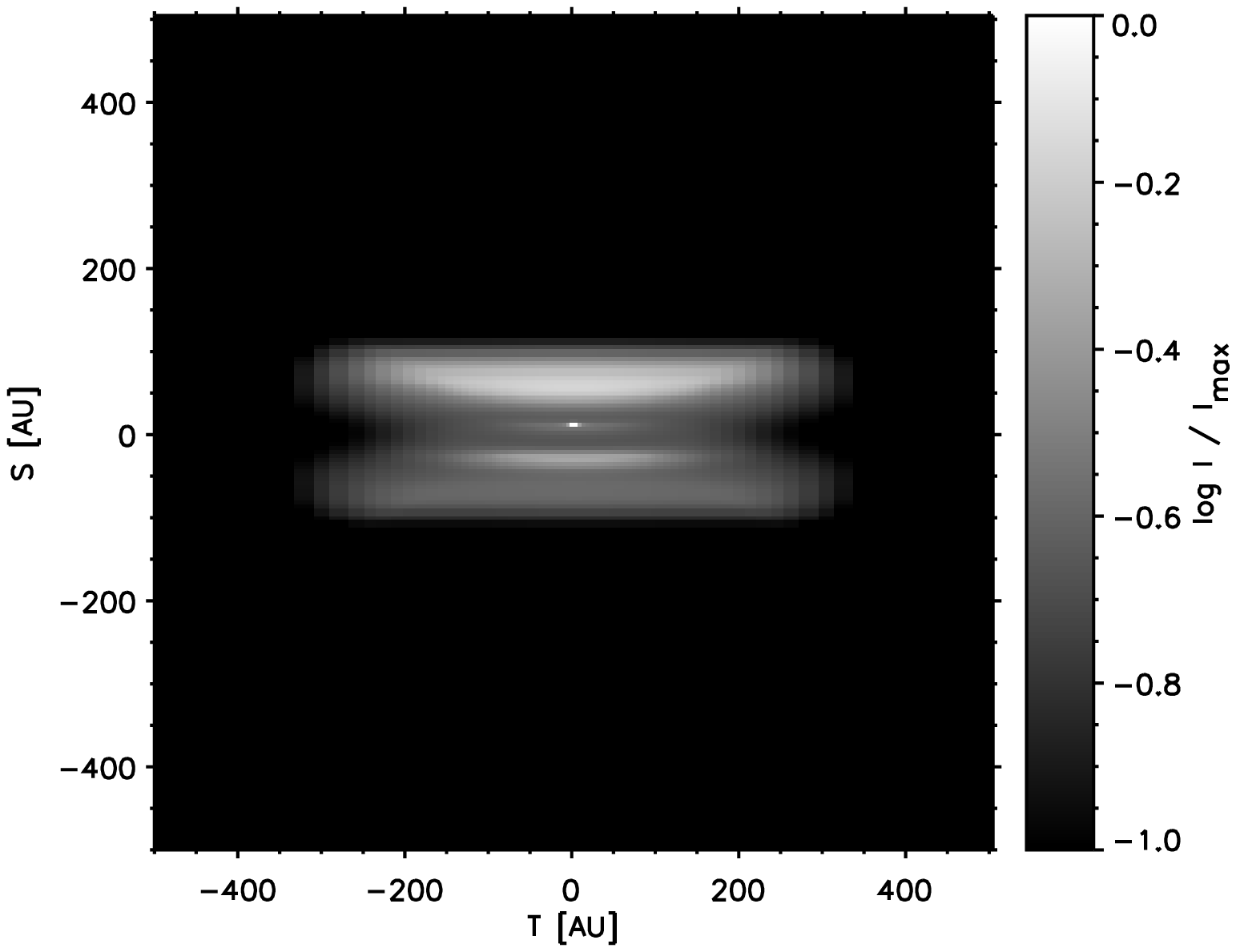}
         \hfil       \includegraphics*[width=80mm]{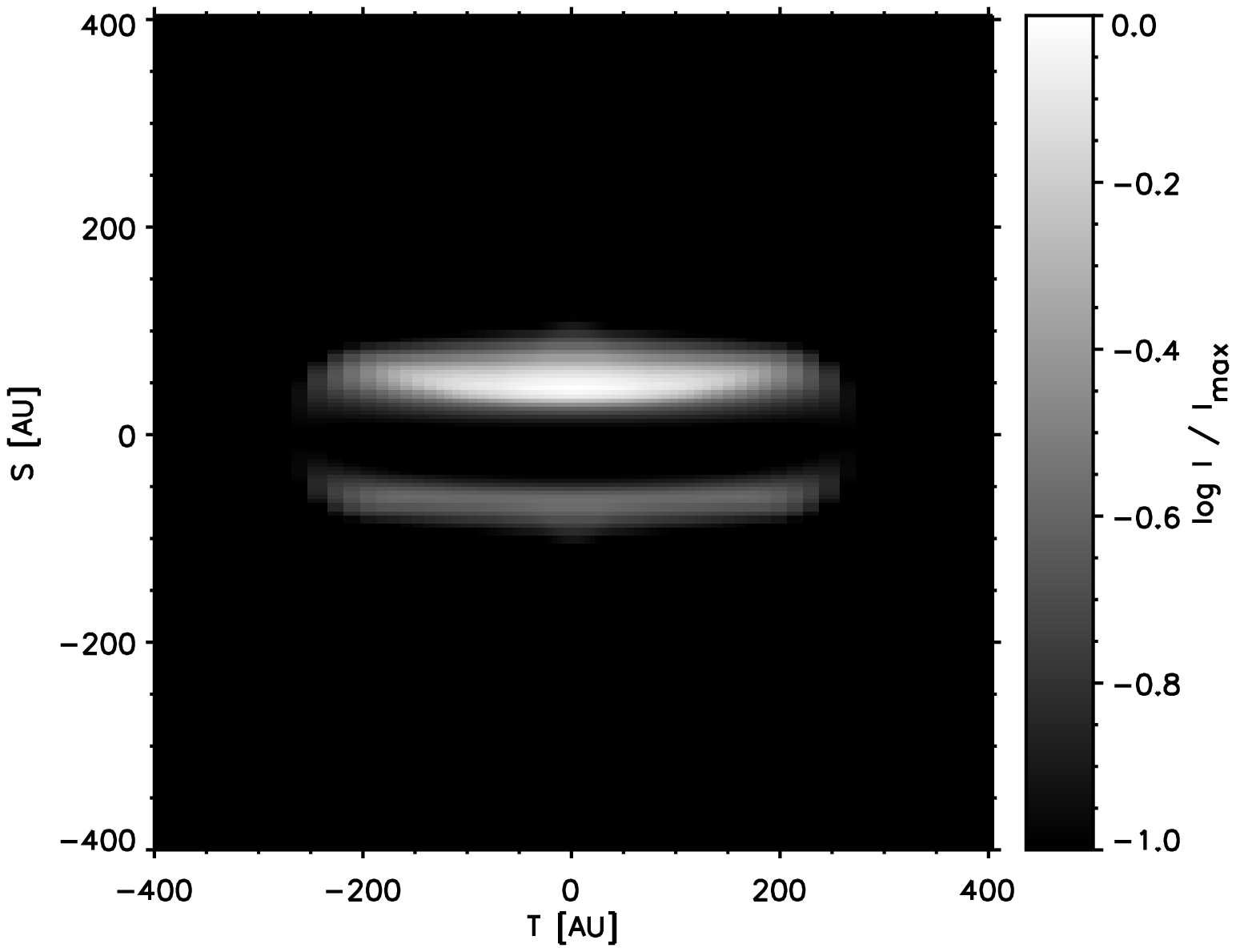}}
\caption{Dust emission of model 3MS\_CCA, 12600~yr. The rotation axis
  is inclined at an angle $\theta=85^\circ$ with respect to the
  line of sight.
  Left: Continuum emission map at $\lambda=3.6\,\mu$m using coagulated dust.
  Right: Emission map assuming unmodified dust.}
\label{PIemcca}
\end{figure*}

In order to compare our numerical results to observations of young
protostellar accretion disks we have produced emission maps and
spectra calculated with a `numerical telescope', which includes the
contribution of scattered light (Yorke \& Bodenheimer 1999).
Figure \ref{PIemsphere} (left panel) shows the dust continuum emission at
3.6\,$\mu$m at the final stage of the collapse of the ``compact sphere'' dust
model (3MS). A dark bar across the midplane of the accretion disk marks a
region of high extinction. Above and below the disk scattered light
reveals the presence of 
the central protostellar radiation source. The spectral energy distribution
(SED) shown in Figure \ref{PIemsphereccaspec} (left panel) displays
well known features of young
dusty protostellar cores (e.g. Sonnhalter, Preibisch \& Yorke 1995): No direct
radiation from the protostar (at an angle of 85$^\circ$), a silicate
absorption feature at $\lambda\approx 10\,\mu$m, and a dust emission
temperature of about 100\,K. For comparison, the corresponding SED
was recalculated for dust with an MRN mass distribution
using the same density distribution. No obvious differences
to the results using the more detailed coagulated dust model
are detectable. This can be attributed to the fact that the
strongly coagulated dust particles are
embedded deeply within the accretion disk, whereas
the outer layers contain only slightly modified
grains. However, when all dust grains are assumed to have a radius of
$a=22\,\mu$m (which corresponds to the maximum grain size formed in model
3MS), the calculated SED is markedly different from that resulting from
dust with an MRN mass distribution.  The
silicate absorption feature is totally absent and more near infrared
emission reaches the observer.

For comparison, an emission map using the unmodified MRN dust distribution
has been calculated and displayed in Figure \ref{PIemsphere} (right panel).
The differences are not overwhelming, but some general tendencies can be
ascertained: For unmodified dust the dark absorption bar across the disk
is somewhat more prominant (especially towards the edges of the disk) and
more scattered light above and below the disk is visible. Because almost
no coagulation occurred in these outer disk regions over the time period
investigated, these differences result primarily from the differential
advection of dust grains.

A similar tendency can be seen for the fractal grains. In Figure
\ref{PIemcca} (left panel) the continuum emission for simulation 3MS\_CCA is
displayed. In contrast to model 3MS the dark absorption bar across the disk is
far more transparent at $\lambda=3.6\,\mu$m when dust coagulation is
permitted. When compared to the emission map resulting from uncoagulated
dust with an MRN mass distribution (Fig. \ref{PIemcca}, right panel), 
it becomes apparant that the coagulation process has enabled the disk to
become rather transparent. The overall disk features are in
general similar to those resulting from the simple spherical grain model.

The SED shown in Figure \ref{PIemsphereccaspec}
(right panel) is similar to the SED of model 3MS (left panel).
The SED generated using the coagulated dust from the 3MS\_CCA simulation
({\it solid line}, right panel)
displays some differences with respect to the SED using
non-coagulated dust: There is a slight shift in the emission peak to
shorter wavelengths, lower far infrared fluxes, enhanced 
mid-infrared emission,
and a less prominent silicate absorption feature.

\section{Discussion and Conclusions}

We have shown that dust coagulation proceeds at an early phase during
the formation of a protostellar accretion disk. Small grains with
$a \la 0.1\,\mu$m are removed from the mass spectrum quickly and
effectively in the midplane of the accretion disk within 
$\sim$10$^3$~yr. Here, large particles with sizes of several
10\,$\mu$m can be produced by coagulation.  The maximum grain size
which can be quickly produced by coagulation
during the collapse and initial accretion of material onto the
disk depends crucially on the assumed sticking strength of the
dust particles. In this respect the process of ice sublimation
plays an important role: When the grain
surface is coated with material which
enhances the grain to grain adhesion, the degree of coagulation 
can be significantly increased.

In the accretion shock front relative velocities of several km\,s$^{-1}$
are achieved due to the size dependent coupling to the gas.
Compact spherical grains decouple at higher gas densities (and
thus earlier during the evolution) than fractal dust
coagulates. Particle shattering of compact spherical grains
was not critically important
during the evolution of the intermediate mass protostars considered
here. We infer, however, by appropriate scaling of masses and
luminosities (see Suttner et al. 1999), that shattering should be
important for high mass protostars.  For
the high mass case radiative acceleration will become increasingly
more effective in causing a size-dependent spread of dust drift 
velocities. Assuming BCCA grains which break apart
at even relatively small collision energies (as in the model of Dominik \&
Tielens 1997), particle shattering gains some importance for the
lower cloud masses considered here. Within the accretion shock grains
are shattered and the maximum grain size is limited to several
$10\,\mu$m. However, the amount of very small debris particles thus
produced is negligible in the total grain mass spectrum.

Gas--dust drift leads to depletion of dust in the immediate
vicinity of the accretion disk everywhere except in the equatorial
regions.  In particular, the gas to dust mass ratio can
be lowered by a factor of 2 to 4 within the accretion shock.
Whereas for a cloud clump mass of 1\,M$_\odot$ 
radiative acceleration of dust grains is negligible, 
for clump masses $\ga 3\,$M$_\odot$ radiative acceleration
of dust grains becomes increasingly important. Depending on how well the
radiation field of the central source is shielded by the disk, the
infall of dust particles can be hindered in the polar regions, whereas
in the equatorial regions the dust moves radially inwards
faster than the gas.

The optical and physical properties of grains are strongly affected
by coagulation.  The specific extinction coefficient in the visual to UV
can be lowered by more than an order
of magnitude in the equatorial plane due to coagulation.  The grain
temperature in the midplane and the grains' capacity for ``freezing
out'' molecules is correspondingly affected.  Although the local
variations of the optical coefficients are large, the only significant
effect to observational properties is a reduction of the near infrared dust
opacity in the wavelength range between 1 and 100\,$\mu$m, which is most
prominent for ``robust'' BCCA particles. Polarization of starlight
should supply an additional appropriate observational tool to determine
the degree of coagulation.

The differences of the global characteristics of the 
simulations using the simple approximation of compact
spherical grains, BPCA dust, and BCCA dust are not as dramatic
as may have been naively expected. For all three cases the hydrodynamical
structure (in particular, gas density and velocity) is strikingly similar.
Thus, we feel justified in using our rather crude dust models to perform
hydrodynamic simulations of low and intermediate mass collapsing
clouds and subsequently assume more sophisticated detailed dust models
to generate emission maps, polarization maps, and SEDs.

Finally, we note that D'Alessio et al. (1999) find that
synthetic 1\,$\mu$m images
of accretion disks around low mass stars appear to have too large
geometrical thicknesses to be consistent with observation, under the
assumption that dust is well mixed with the gas.  Our study shows that
the issue might be resolved by taking into proper account the differential
advection of dust grains.

\acknowledgements
We are grateful to Thomas Henning, Doug Lin, and Rainer Schr\"apler for
helpful discussions and to an anonymous referee for useful suggestions.
The research described in this paper was carried
out by the Jet Propulsion Laboratory (JPL), California Institute of
Technology, and was supported by the ``Deutsche Forschungsgemeinschaft''
(DFG) under the ``Physics of Star Formation'' program (grant Yo 5/20-2)
and the National Aeronautics and Space Administration (NASA) under grant
NRA-99-01-ATP-065.  The calculations were performed on workstations at
JPL and the ``Rechenzentrum der Universit\"at W\"urzburg'', on a Cray
T90 at the ``HLRZ J\"ulich'' and on a
SP2 parallel computer at the same facility.

\appendix
\begin{center}
{\bf APPENDIX}
\end{center}
\begin{eqsecnum}

Here, we give an analytic expression for the kernel
$\gamma(m,m^{'},m^{''}\!\!,\delta v(m^{'}, m^{''}))$ in the shattering equation
of section \ref{SEcoagshat}. As stated in section \ref{SEcoagshatsol}, the
integral of the shattering equation is discretized by summing over the
dust mass space ranging from $m_1$ to $m_N$ (or,
equivalently, from $a_1$ to $a_N$).
Thus, $\gamma$ transforms to the discrete function $g_{ijk}$,
from which the debris distribution
$G_k(m_i,m_j,\delta v_{ij})=g_{ijk} m_k/(m_i+m_j)$ can be separated.  \\

\noindent
For $v_{\rm crit} \le \delta v_{i,j} < v_{\rm cat}$ and
$w m_j\in [m_k^- , m_k^+] $:
\begin{eqnarray}\label{eq-G1}
  G_k(m_i,m_j,\delta v_{ij}) = \frac{m_j}{m_i+m_j}(1-w) 
\end{eqnarray}
For $v_{\rm crit} \le \delta v_{i,j} < v_{\rm cat}$ and $a_k \le a_{\rm max}$:
\begin{eqnarray}\label{eq-G2}
  G_k(m_i,m_j,\delta v_{ij}) = \frac{m_j}{m_i+m_j}w\ f_{\rm MRN}
\end{eqnarray}
For $\delta v_{i,j} \ge v_{\rm cat}$ and $a_k \le a_{\rm max}$:
\begin{eqnarray}\label{eq-G3}
  G_k(m_i,m_j,\delta v_{ij}) = f_{\rm MRN}
\end{eqnarray}
Otherwise,  $G_k(m_i,m_j,\delta v_{ij}) = 0$.

We have used the following assumptions and definitions:
\begin{eqnarray*}
  m_i &\ge& m_j                                             \\
  m_k^- &=& (m_k + m_{k-1}) / 2 \quad m_k^+ = m_{k+1}^-     \\
  w   &=& \left(\frac{\delta v_{ij}}
          {3.64\ \rm km\ s^{-1}}\right)^{16/9}              \\
  v_{\rm crit}&=&2.7\ {\rm km\ s^{-1}}                      \\
  v_{\rm cat} &=& {\rm max}
          \left(v_{\rm crit},\ 1.13\ {\rm km\ s^{-1}}
          \left(m_i / m_j \right)^{9/16}\right)             \\
  a_{\rm max}&=&\cases{
          0.28\ (w\ m_j / \varrho_{\rm bulk} )^{1/3}
          &if $\delta v_{ij} < v_{\rm cat}$             \cr
          0.20\ a_i\ v_{\rm cat} / \delta v_{ij}
          &if $\delta v_{ij} \ge v_{\rm cat}$           \cr
}
\end{eqnarray*}

The formulae were adapted from the work of Jones et al. (1996). Here,
$w$ denotes the ejected crater mass in units of the projectile
mass, $v_{\rm cat}$ the critical velocity for the onset of total
disruption of the target and $a_{\rm max}$ the radius of the
largest debris particle. The debris
particles are redistributed according to a MRN size distribution
$f_{\rm MRN}$ between $a_{\rm min}=a_1$ and $a_{\rm max}$
(i.e. between $m_{\rm min}=m_1$ and $m_{\rm max}$):

\begin{eqnarray}
f_{\rm MRN}&=&\frac{m_k^{-5/6} \Delta m_k}{\sum_{i=1}^{\rm max}
  m_i^{-5/6}\Delta m_i}
\end{eqnarray}
\end{eqsecnum}

\end{document}